\documentclass[10pt]{amsart}
\usepackage{amsmath}
\usepackage{graphics, graphicx}
\usepackage[english,  activeacute]{babel}
\usepackage[latin1]{inputenc}
\usepackage{amssymb}
\usepackage{amsthm}
\usepackage{array}
\usepackage{a4wide}
\usepackage{color, url}
\usepackage{floatrow}
\usepackage{pstricks-add}

\newcommand{\mat}[2]{\left[ \begin{array}{*{#1}{c}}#2\end{array}\right]}

\begin{document}

\thispagestyle{empty}
\setlength{\parindent}{11pt}

\title[An algebraic approach to electron interactions]{An algebraic approach to electron interactions in quantum Hall systems}

\author{S. B. Mulay}
\address{\noindent Department of Mathematics, University of Tennessee, Knoxville, TN 37996, USA}

\author{J. J. Quinn}
\address{\noindent Department of Physics, University of Tennessee, Knoxville, TN 37996, USA}

\author{M. A. Shattuck}
\address{\noindent Department of Mathematics, University of Tennessee, Knoxville, TN 37996, USA}

\date{\today}

\maketitle
\noindent \emph{Abstract}\\

Let $m$ denote the number of quasielectrons (QEs) in a quantum Hall system containing $N$ particles altogether.  We show in several general cases that for systems containing $m$ QEs in a single angular momentum shell above $N-m$ Fermions in an incompressible quantum liquid (IQL) state having filling factor $\nu={^1/_3}$ that there always exists a configuration whose symmetric correlation function $G$ is nonzero.  This extends recent comparable results concerning the IQL state.  As a consequence, one can obtain (explicitly) a configuration having a nonzero $G$ for all $N \leq 8$ particle systems containing any number of QEs. To establish our result, we construct a family of multi-graphs on $N$ vertices satisfying certain restraints on the degrees of the vertices and possessing the property that whenever one computes the linear symmetrization of the graph monomial of any member of the family, the result is always nonzero.  The nonzero linear symmetrization that is obtained in each case is in fact an example of what is called a \emph{relative semi-invariant} of a (generic) binary form of degree $N$.  Thus, in addition to providing new correlation functions for systems of interacting Fermions containing QEs, our construction could be of interest from both the invariant and graph theoretic standpoints.\\

\section{Introduction}

A trial wave function $\Psi(z_1,\ldots,z_N)$ of an $N$ electron system can always be expressed as the product of an
antisymmetric Fermion factor $F = \prod_{1 \leq i < j \leq N}(z_i-z_j)$, and a symmetric correlation factor $G=G(z_1,\ldots,z_N)$ that takes into account Coulomb interactions.   In this paper, we will address certain mathematical aspects of the latter.  Let $z_{ij} = z_i - z_j$, where $z_i$ is the complex coordinate of the $i^{\rm{th}}$ electron, though here we will regard each $z_i$ as an indeterminate. We refer to $z_{ij}$ as a correlation factor (cf), even when it arises from the Pauli principle. One may take $G = 1$ for systems of non-interacting Fermions.  It will be convenient to represent the Coulomb interactions diagrammatically as a multi-graph on $N$ vertices where the edges denote cf factors.

For example, in the incompressible quantum liquid (IQL) state at filling factor $\nu = {^1/_3}$ (see \cite{Laughlin83}), two cf lines connect each pair of Fermions.  For each labeling of the vertices of the corresponding multi-graph, one takes the product of all the cf factors, and then computes the sum of the products corresponding to all possible labelings to obtain the correlation $G$. Another example involves the Moore-Read state \cite{Moore91} of the half filled first excited Landau level (LL1) with $\nu = 2 + {^1/_2}$, where $N$ is even and the $N$ electrons for LL1 are partitioned into two subsets $A$ and $B$, each of size $m=N/2$, with two cfs joining each pair of electrons in $A$ and also each pair in $B$, as noted in \cite{Quinn14}. To compute the correlation $G$ in this case, one takes the product of all cf factors in the diagram corresponding to a given partition $(A,B)$, and then sums these products over all possible partitions to obtain $G$.  In general, in order for a given configuration to exist,  it is necessary that the symmetric correlation function $G$ work out to be nonzero. Otherwise, the configuration is said to be non-existent.  In both of the Laughlin and Moore-Read cases described above, it can be shown that $G$ is nonzero and hence the associated configurations are always existent.

Jain \cite{Jain89, Jain90} introduced a more general composite Fermion (CF) picture that correctly predicts the IQL states at filling factors $\nu=\frac{n}{2pn\pm1}$, where $n$ and $p$ are positive integers, which correspond to integrally filled CF Landau levels. The Jain-Laughlin sequence of mean field CF states is the most robust set of fractional quantum Hall states observed experimentally.  Making use of Haldane's spherical geometry~\cite{Haldane83, Haldane85}, Chen and Quinn~\cite{Chen94} introduced an \textquotedblleft
effective CF angular momentum\textquotedblright\space$l_0^\ast=l-p(N-1)$ associated with the lowest CF Landau level (CFLL0).
For $N = 2l_0^\ast+1$, this level is exactly filled and a Jain IQL state results. If $N>2l_0^\ast+1$, then $N-(2l_0^\ast+1)$
particles must be placed in the next angular momentum shell with $l_{\text{QE}}=l_1^\ast = l_0^\ast+1$; these are CF quasielectrons (QEs).
If $N<2l_0^\ast+1$, there will be $2l_0^\ast+1-N$ CF quasiholes (QHs) in CFLL0, with
$l_{\text{QH}} = l_0^\ast$. For any given value of $l$, the single electron
angular momentum, one can obtain the number of QEs in the partially filled shell (or the number of QHs in the partially
unfilled shell). The lowest band of angular momentum states will contain the minimum number of CF quasiparticle excitations consistent
with the values of $2l$ and $N$. The value of $(2l,N)$ defines the function space of the $N$ electron system.

The correlation factor $G$ must then satisfy a number of conditions. For example, the highest power of $z_i$ in any term of
$G$ cannot exceed $2l + 1 - N$. In addition, the value of the total angular momentum of the correlated state must satisfy
the equation $L = \left(N/2\right)\,\left(2l+1-N\right)- \kappa_G$, where $\kappa _G$ is the degree of the homogeneous polynomial
$G$. Knowing the value of $L$ for IQL states and for states containing a few quasielectrons (or a few quasiholes) from Jain's
mean field CF picture allows one to determine $\kappa _G$.

Laughlin~\cite{Laughlin83} realized that if the interacting electrons could avoid the most strongly repulsive pair states, an incompressible quantum liquid state could result.
He suggested a trial wave function for a filling factor $\nu$ equal to the reciprocal of an odd positive integer $m$, in which the correlation function, denoted by $G_L$, was given by $\prod_{1\leq i<j\leq N}{(z_i-z_j)^{m-1}}$.
One can represent the configuration for $G_L$ diagrammatically by distributing $N$ dots along the circumference of a circle, denoting $N$ electrons, and drawing $\frac{m-1}{2}$ double lines between every pair of electrons, each denoting two cfs.  Note that $G_L$ is nonzero, being an integral power of the discriminant of the $z_i$, whence the configuration is existent.  In a previous paper \cite[Theorem 1]{MQS}, it was shown, more generally, that in fact there exist configurations of $N$ Fermions in the IQL state having a nonzero correlation function $G$ for all filling factors of the form $\frac{n}{2pn\pm1}<{^1/_2}$ where $N$ is assumed to be a multiple of $n$ (the $n=1$ case corresponding to Laughlin).

Here, we wish to extend these results to systems containing quasielectrons.  More specifically, we identify existent configurations for a system containing $m$ QEs in a single momentum shell above $N-m$ Fermions in an IQL state having filling factor $\nu={^1/_3}$.  Combining general results with some specific computations covers all cases where $N \leq 8$.  To obtain the configurations, we construct a family of undirected, loopless multi-graphs on $N$ vertices whose (reciprocal) graph monomials when symmetrized are nonzero (see Theorem 3 below and Corollary).  Particularizing our results to quantum Hall systems of $N$ Fermions containing $m$ QEs as described yields the following.\\

\noindent \underline {\bf Theorem 1}: Let $N\geq 3$ be a positive integer and $m$ be an integer belonging to $\{1,2,N/2,(N+1)/2,N/2+1\}$.  For systems containing $m$ QEs in a single momentum shell above $N-m$ Fermions in an IQL state having filling factor $\nu={^1/_3}$, it is always possible to find a configuration having nonzero symmetric correlation function $G$.  In particular, for a system  containing $N \leq 8$ Fermions in all and any number of QEs, one can always find such a configuration.\\

\noindent Note that the inherent difficulty of generalizing Theorem 1 to any number of QEs lies in the explicit computation of all possible values of the total angular momentum $L$; for unlike in the IQL case, $L$ can assume several positive values.  Moreover, even having determined the set of all possible values of $L$, a major hurdle still lies in finding existent configurations of the required type for each $L$.

The organization of this paper is as follows.  In section II, we discuss the algebra of correlation functions and formulate the problem in terms of multi-graphs, recalling some standard terminology.  We present in section III our main results featuring the construction of certain kinds of \text{(semi-)} invariants.  In the final section, we discuss applications of our results to quantum systems containing QEs as described above and show how Theorem 1 follows as a consequence.

\section{Preliminaries}

Recall that
a correlation diagram for $N$ Fermions graphically exhibits the potencies
of their mutual interactions and so, in purely mathematical terms, it is
a (undirected, loopless) {\it multi-graph} on $N$ vertices. The prefix
{\it multi-} is indicative of the fact that some of the vertex-pairs may be connected by more
than one edge. Here, we will regard {\it correlation diagram} and {\it multi-graph}
as equivalent terms. Given a multi-graph on $N$ vertices, a choice of a labeling
of its vertices by the numbers $1, 2, \dots, N$ gives rise to a product of terms
$(z_{i} - z_{j})^{p_{ij}}$, where $z_{i}$ is an indeterminate for $1 \leq i \leq N$ and
the nonnegative integer $p_{ij}$ for $1 \leq i < j \leq N$ is the number of edges between
the vertices labeled $i$ and $j$ in the multi-graph. In the classical theory of invariants,
a product of this type is known as a {\it graph-monomial} (see, e.g., \cite{AS}). Note that since our $N$ Fermions are
indistinguishable, we must consider each of the possible choices of vertex-labelings, for the
correlation diagram under consideration, on an equal footing. Commonly, two multi-graphs
on $N$ vertices are called {\it isomorphic} if one is obtained from the other by a
relabeling of its vertices (see Figure 1 below for an example of isomorphic multi-graphs).

\begin{figure}[h!]
\centering
\begin{tabular}{cc}
\includegraphics[scale=0.3]{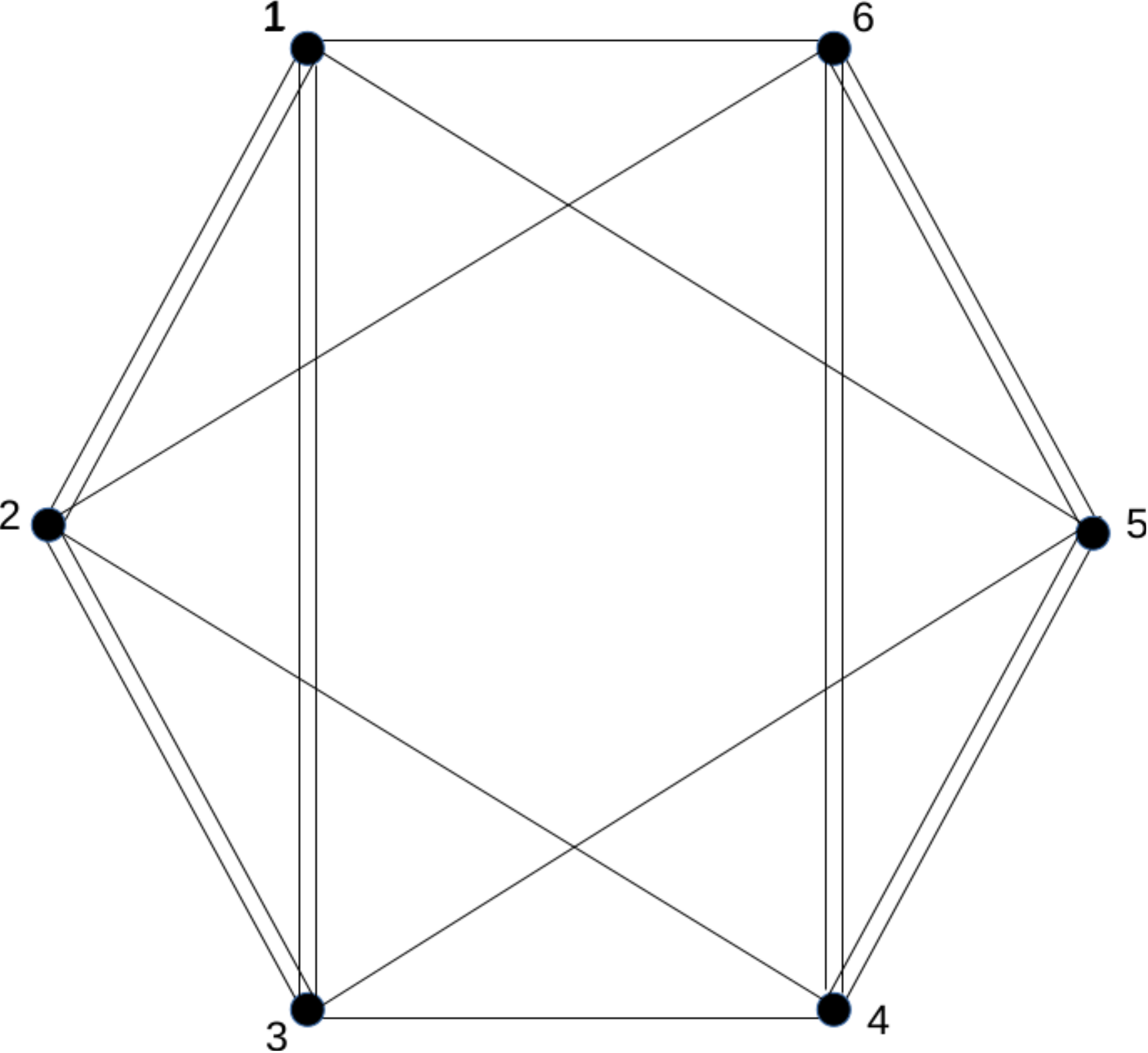} & \includegraphics[scale=0.3]{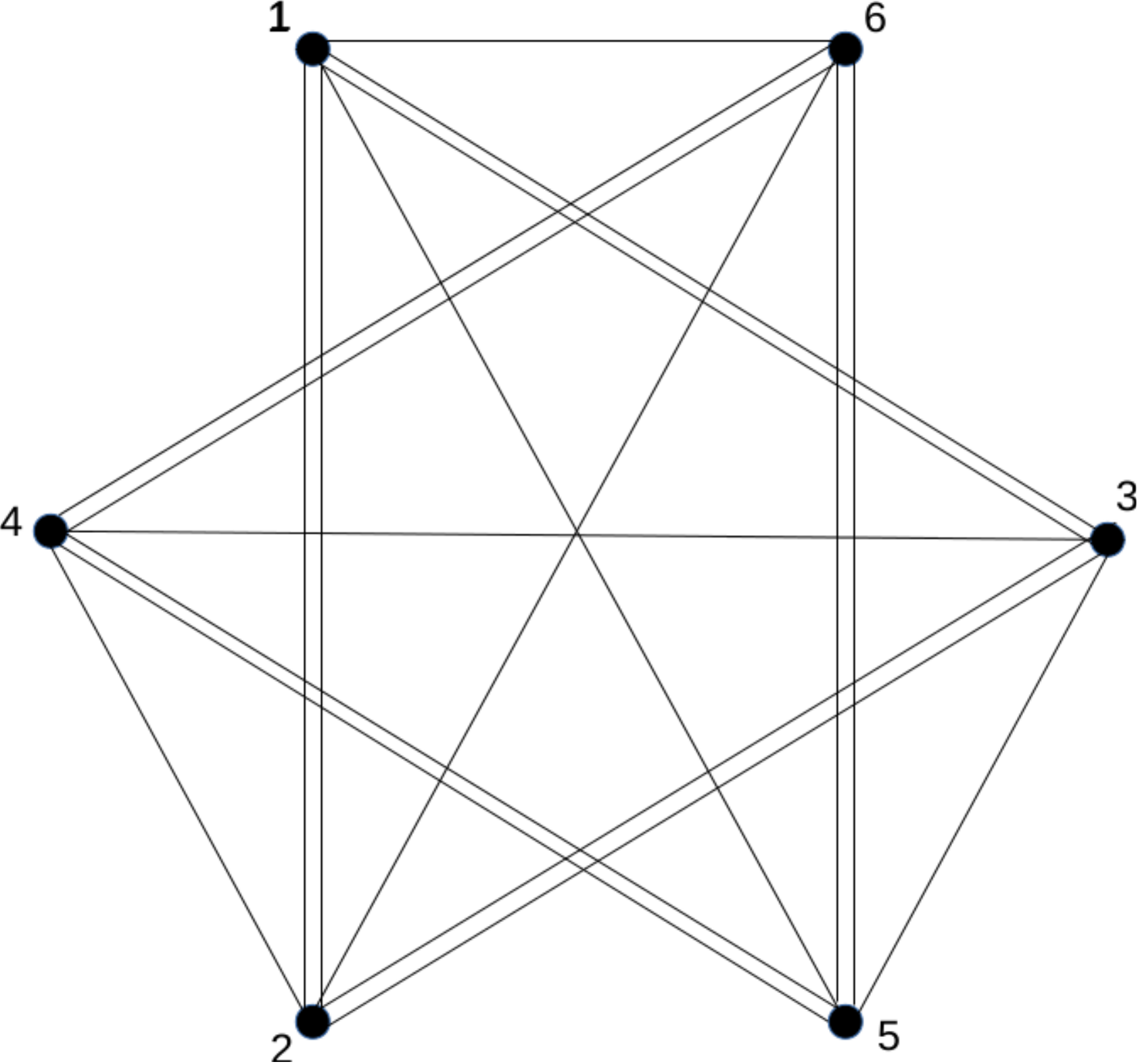}
\end{tabular}
\caption{\small Isomorphic multi-graphs on $6$ vertices.}
\end{figure}

The isomorphism class of a correlation diagram may be regarded as a
{\it configuration} of interacting Fermions; nonisomorphic correlation diagrams
correspond to distinct configurations. The correlation function of a
configuration of $N$ interacting Fermions is defined to be the sum (if
preferred, it can also be defined as the average) of the graph-monomials associated to
that configuration. In other words, if we pick one correlation diagram for the
configuration and call its associated graph-monomial $f(z_{1}, \dots, z_{N})$, then
the correlation function of the configuration is the {\it symmetrization} of $f$,
{\it i.e.}, $\sum f(z_{\sigma (1)}, \dots , z_{\sigma (N)})$, where the sum ranges
over all permutations $\sigma$ of $\{1,2,\dots, N \}$. Clearly, such a correlation
function is a homogeneous polynomial symmetric in $z_{1},  \dots , z_{N}$. We deem the configuration as {\em existent} when this correlation function is not identically zero, and as
{\em nonexistent} otherwise. If correlation functions of two configurations are the same
up to a nonzero numerical (rational) factor, then the configurations are regarded as
{\it equivalent}.

It is worth noting that on account of the symmetries inherent to a given multi-graph, it
can very well be the case that certain distinct labelings of vertices yield the same
graph-monomial. From a computational point of view, the correlation function of a configuration
is easier to deal with when its corresponding set of graph-monomials is small and hence
multi-graphs with many intrinsic symmetries are perhaps more desirable. In the extremal example of a
multi-graph where the number of edges between any two vertices is the same integer $e$
({\it i.e.}, $p_{ij} = e$ for $1 \leq i < j \leq N$), there are at most two distinct
graph-monomials for the configuration (differing only by a factor of $\pm 1$). Recall that
such is precisely the case if one were to consider the Laughlin configuration for the IQL state with filling
factor $\nu = 1 / (2 p + 1)$ (forcing $p_{ij} = e = 2 p$). In general, a simple exercise shows that
the graph-monomial of a multi-graph on $N$ vertices is a symmetric polynomial in the variables
$z_{1}, \dots , z_{N}$ if and only if there is an integer $p$ such that $p_{ij} = 2 p$ for
all $1 \leq i < j \leq N$.

For a system of $N$ interacting Fermions, their individual angular momenta, together with
the filling factor $\nu$, dictate an upper bound $d$ on the degree of a vertex ({\it i.e.}, the
number of edges emanating from a vertex) in the corresponding correlation diagram, whereas the
total angular momentum $L$ of the system demands that the corresponding correlation function
be a homogeneous polynomial of (total) degree $(N d / 2) - L$. Usually, there are several possible
configurations that meet these dictated requirements; their number increases rather sharply with
increasing values of $N$. To determine which of these configurations actually exist, it is
essential to ascertain the nonzero-ness of their corresponding correlation functions. This is a
nontrivial task when the associated correlation diagram has vertex-pairs connected by an odd number
of edges. Even more challenging is the problem of determining, in some concrete manner, the set of
equivalence classes of these configurations.

In general, if $L = 0$, then it turns out that each vertex
in a related correlation diagram must have the same maximum allowed degree $d$. A (undirected, loopless)
multi-graph each of whose vertices has the same degree $d$ is said to be {\it $d$-regular}. The problem of
counting  distinct configurations of $N$ Fermions with $L = 0$ and a given filling factor $\nu$
translates to counting isomorphism classes of $d$-regular loopless multi-graphs on $N$ vertices.
We point out that this counting problem appears to be largely open and is a subject of ongoing research
(see~\cite{GM}).  When $L>0$, then some of the vertices will fail to have the maximum allowed degree and the problem translates into determining classes of loopless multi-graphs on $N$ vertices in which there is a common upper bound on the degree of each vertex. Here, we will be interested in determining the nonzero-ness of the associated correlation functions in some particular cases when $L>0$. For comparable results when $L=0$, see \cite{MQS}.

\begin{table}[h!]
\label{tab:1}
\begin{tabular}{|c|c|c|c|c|c|}
\hline
$l$	  &$l_0^\ast$&$n_{\text{QE}}$&$l_{\text{QE}}$&$k_M$&$L$ \\ \hline
 4.5  & 1.5 &  0  & 2.5 & 6   & 0   \\ \hline
 4    & 1   &  1  & 2   & 5   & 2   \\ \hline
 3.5  & 0.5 &  2  & 1.5 & 4   & 0$\oplus$2   \\ \hline
 3    & 0   &  3  & 1   & 3   & 0   \\ \hline
\end{tabular}\medskip
\caption{\small Values of $l$ for an $N=4$ electron system and the values of $l_0^\ast$, $n_{\text{QE}}$,
$l_{\text{QE}}$, $k_M$, and $L$ which result.}
\end{table}

Let us now consider the situation with $N=4$ electron systems having $\nu = {^1/_3}$ filled IQL states
and its excited states containing one, two, and three QEs. These correspond to $2l$ values of $9$, $8$, $7$, and $6$, respectively, where $l$ denotes the single electron angular momentum.  In Table 1 above, we summarize the results of Jain's mean field CF picture \cite{Jain89, Jain90} applied to the various low energy states. The table gives the values of $l$ and the resulting values of
$l^\ast_0 = l - \left(N-1\right)$, the CF angular momentum; $n_{\text{QE}} = N - \left(2l^\ast_0+1\right)$,
the number of QEs; $l_{\text{QE}}$, the QE angular momentum; $k_M = 2l-\left(N-1\right)$, the maximum number of
correlation factor (cf) lines that can emanate from an electron in the correlation diagram; and the allowed
values of the total angular momentum $L$ which result.  Concerning the question of existence of configurations, one would need to construct in each of four cases a loopless graph whose graph monomial is nonzero when symmetrized in which the degree of every vertex is bounded above by $k_M$, with half the sum of all the degrees given by $(N/2)(2l+1-N)-L$.  Note that $L$ is in general not uniquely determined by the number of QEs, as witnessed here.

\begin{figure}[!h]
\centering
\begin{floatrow}
\ffigbox{\includegraphics[scale=0.4]{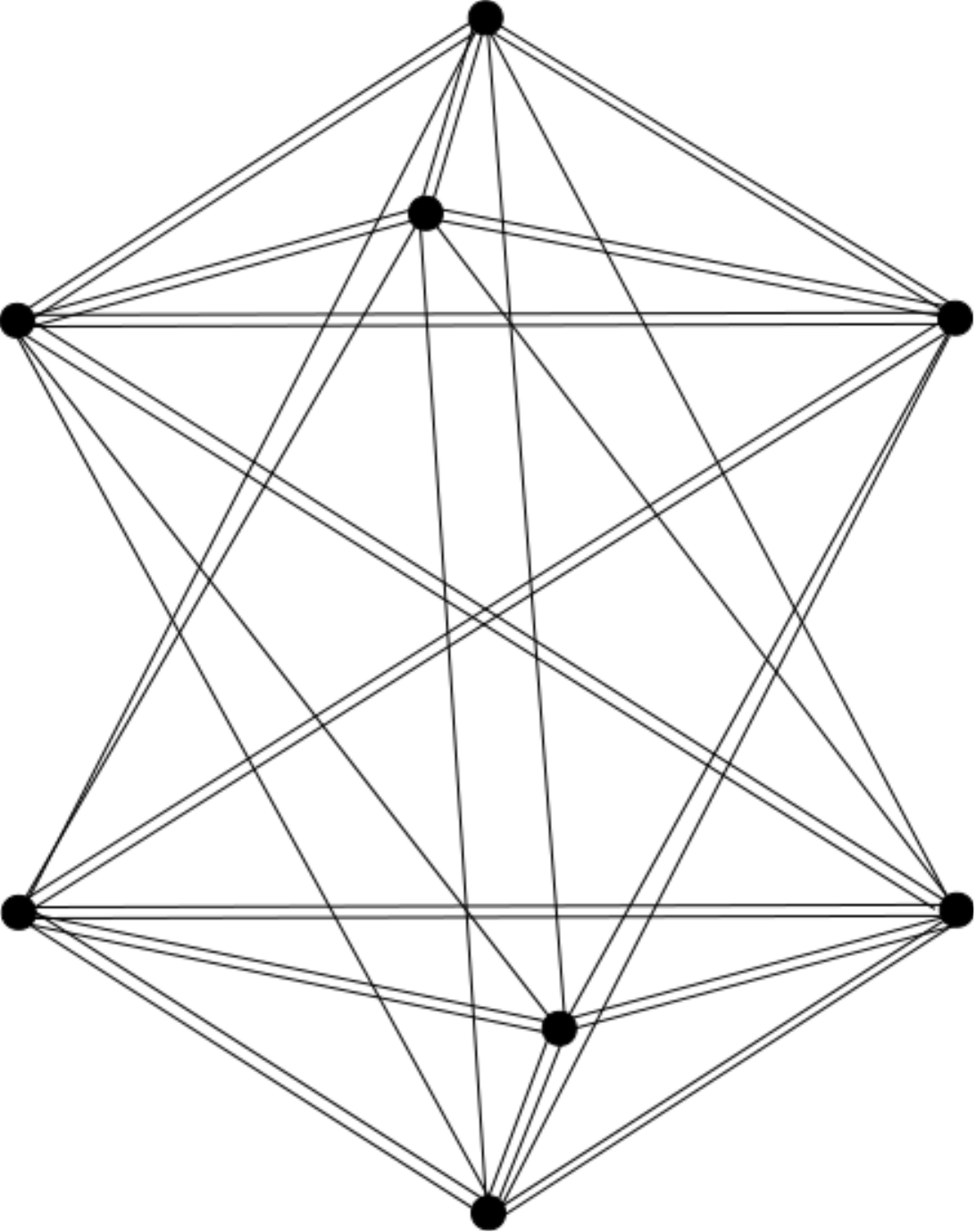}}{\caption{\small $(N, QE, \nu, L) = (8, 4, 1/3, 2)$.}}
\ffigbox{\includegraphics[scale=0.4]{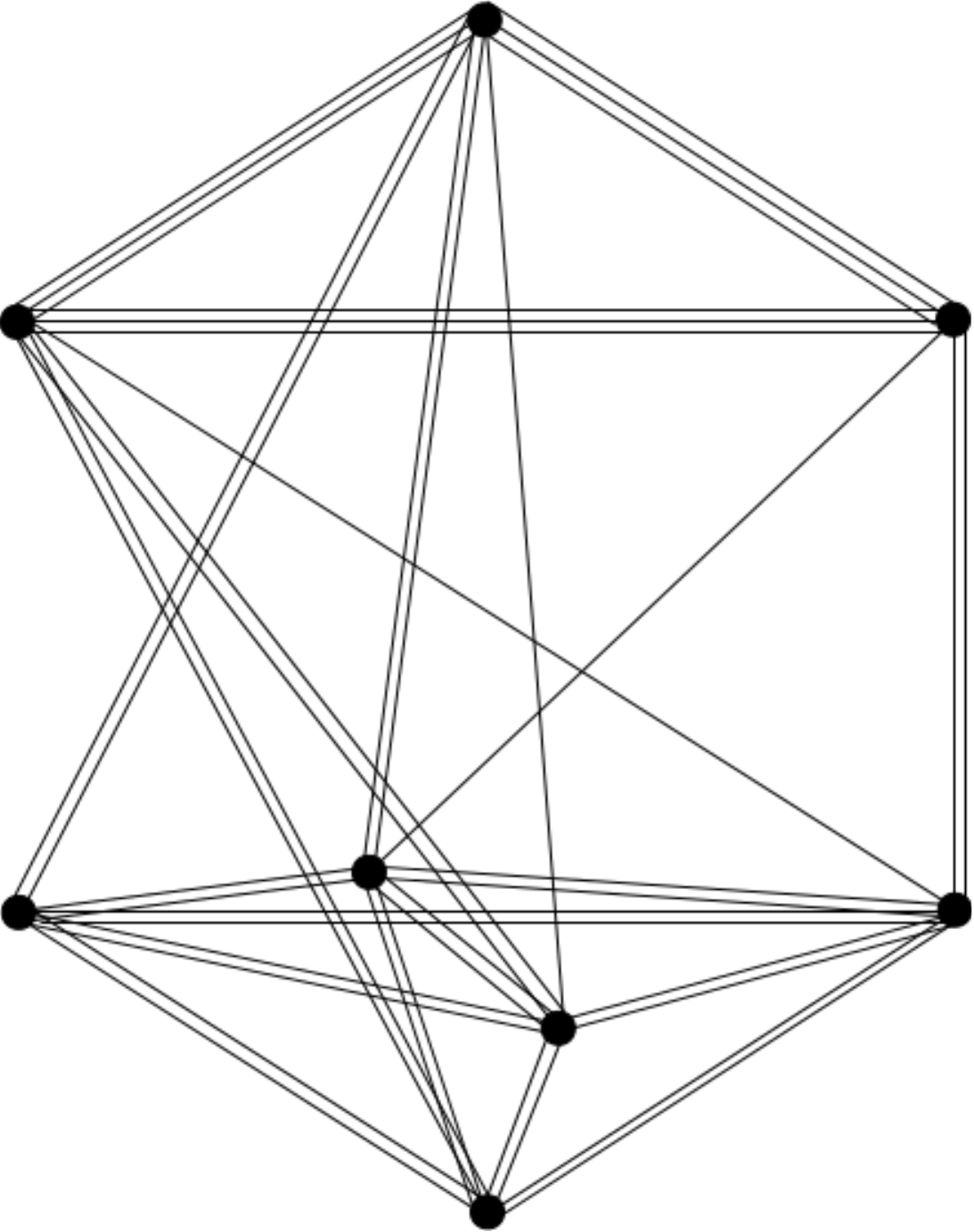}}{\caption{\small $(N, QE, \nu, L) = (8, 3, 1/3, 2)$.}}
\end{floatrow}
\end{figure}

Consider an IQL state with $N$ particles where $\nu=\frac{1}{2p+1}$ and an excited state containing $m=N-2l^\ast_0-1$ QEs where $l^\ast_0=l-p(N-1)$, which implies $m=(2p+1)(N-1)-2l$.  In particular, for $p=1$ (\emph{i.e.}, $\nu={^1/_3}$), we have $k_M=2l-N+1=2(N-1)-m$.  In Figures 2 and 3 below, we illustrate two correlation diagrams corresponding to existent Fermion systems containing $m$ QEs in an excited state above $N-m$ Fermions in an IQL state with $N=8$ and $\nu={^1/_3}$ for $m=4$ and $m=3$, respectively, and $L=2$.  Note that the total number of cf lines in each diagram equals the total degree $\kappa_G=\frac{N(2N-2-m)}{2}-L$, which works out to $38$ and $42$, respectively, as seen in the figures.

It is well-known that the symmetrized graph-monomial of an undirected loopless multi-graph is
also called a {\it relative semi-invariant} of a (generic) binary form of degree $N$. If the multi-graph is $d$-regular, then the associated symmetrized graph-monomial is a {\it
relative invariant} of the degree $N$ binary form.  Ever since Cayley founded the theory of invariants, explicit construction of (semi-) invariants has been of extensive interest. Though our motivation for the explicit constructions of invariants formulated in the next section lies in building correlation functions for systems of interacting Fermions, these results have more to offer from a purely invariant theoretic viewpoint.  For a deeper, more comprehensive treatment of the theory of invariants of binary forms, we refer the interested reader to either the classic~\cite{GY} or the more contemporary exposition~\cite{KR}.

Although multi-graphs can be visually pleasing, it is undoubtedly simpler to deal with their adjacency matrices
in attempting to prove precise results. Thus the reader will find our definitions and theorems formulated in the language of matrices.

We conclude this section by recalling some notation and terminology.  We denote the sets of ordinary integers, nonnegative integers and rational numbers by ${\Bbb Z}$, ${\Bbb N}$ and ${\Bbb Q}$, respectively. For a function $f$
defined on a set $S$, by $f(S)$, we mean the set $\{ f(a) \mid a \in S \}$.  We use the notation $|S|$ to denote the cardinality of $S$. Here, we are mainly interested in polynomials
and rational functions having coefficients in an integral domain of characteristic zero, in particular, in a field containing ${\Bbb Q}$. Since the notions of {\it degree} and {\it order} of a rational function play important roles in our proofs that follow, we remind the reader now of their definitions and basic properties. Consider a rational
function $f$ in a set of indeterminates $z$ such that $f = P / Q$ for some nonzero polynomials $P$ and $Q$
in $z$ having coefficients in an integral domain $k$. Then the degree of $f$ is defined to be the difference
between the (usual) degrees of $P$ and $Q$. By convention, $0$ has degree $ - \infty$. Let $g$ be also
a rational function in $z$ with coefficients in $k$. Recall that the degree of $f g$ is the sum of the
degrees of $f$ and $g$ whereas the degree of $f + g$ is bounded above by the maximum of the degrees of $f$ and
$g$. Moreover, the degree of $f + g$ is the maximum of the degrees of $f$ and $g$ whenever $f$ and $g$ have
unequal degrees. Now suppose $k$ is a unique factorization domain and $J$ is a nonzero principal prime ideal of
the polynomial ring $k [z]$. Then the {\it $J$-order} of a nonzero polynomial $h \in k[z]$ is defined to be the largest
nonnegative integer $m$ such that $h$ is in $J^{m}$. Subsequently, the $J$-order of $f$ is defined to be the
difference between the $J$-orders of $P$ and $Q$. By convention, the $J$-order of $0$ is $\infty$. If $u \in k[z]$
is a generator of $J$, then the term {\it $u$-order} is regarded to be synonymous with the term $J$-order. Recall
that the $J$-order of $f g$ is the sum of their respective $J$-orders whereas the $J$-order of $f + g$ is bounded
below by the minimum of the $J$-orders of $f$ and $g$. Moreover, the $J$-order of $f + g$ equals the minimum of the
$J$-orders of $f$ and $g$ whenever $f$ and $g$ have unequal $J$-orders. For various other notions from basic abstract algebra that are tacitly used below, the reader is referred to \cite{SZ}. \\

\section{Construction of invariants}

\noindent \underline {\bf Definitions}: Let $N\geq2$ be an integer and $k$ be a field containing ${\Bbb Q}$.
Let $z_{1}, \dots, z_{N}$ be indeterminates and $z$ stand for
$(z_{1}, \dots , z_{N})$.
\begin{enumerate}
\item By $S_{N}$, we denote the group of all permutations of $\{1 , \ldots, N \}$.
For any ring $k$, let
\[ Symm_{N} \, : k [z_{1}, \dots , z_{N}] \rightarrow k [z_{1}, \dots , z_{N}] \]
be the {\it Symmetrization operator} given by
\[ Symm_{N} (f(z_{1}, \ldots , z_{N})) :=
\sum_{\sigma \in S_{N}} f(z_{\sigma (1)}, \ldots , z_{\sigma (N)}) .\]
$f$ is {\it symmetric} provided
$f(z_{\sigma (1)}, \ldots , z_{\sigma (N)}) = f(z_{1}, \ldots , z_{N})$ for all
$\sigma \in S_{N}$.
\item Given an $N \times N$ matrix $A := [a_{ij}]$ with integer entries, let
$r_{i}$ denote the sum of the entries in the $i$-th row of $A$ for $1 \leq i \leq N$
and define
\[ \rho (A)\, : = \, (r_{1}, \ldots , r_{N}) \, . \]
\item Given an $N \times N$ matrix $A := [a_{ij}]$, where each $a_{ij}$ is a nonnegative
integer, letting $z$ stand for the vector $(z_{1}, \ldots , z_{N})$, define
\[ \delta (z,\, A) \, := \, \prod _{1 \leq i < j \leq N} \, (z_{i} - z_{j})^{a_{ij}} . \]
\item Let $E (N)$ denote the set of all $N \times N$ symmetric matrices $A:=[a_{ij}]$ such that
each $a_{ij}$ is a nonnegative integer and $a_{ii} = 0$ for $1 \leq i \leq N$. For
$V := (d_{1}, \ldots , d_{N}) \in {\Bbb Z}^N$, let $E(N, \leq V)$ be the subset of $E (N)$
consisting of all $A \in E (N)$ such that letting $\rho(A):=(r_1,\ldots,r_N)$, we have $r_i\leq d_i$ for all $1 \leq i \leq N$.  Let $E(N,V)$ be the subset consisting of $A \in E(N,\leq V)$ such that $\rho (A)=V$.  If $V = (d, d, \ldots , d)$, then the sets $E(N,\leq V)$ and $E(N,V)$ will be denoted respectively by $E(N,\leq d)$ and $E(N,d)$.
\item For a positive integers $m$, $n$, define $D_{(m, n)}$ to be the $m \times n$ matrix
$[c_{ij}]$, where
\[ c_{ij} \, :=  \, \left \{ \begin{array}{ll} 0 & \mbox{if $i = j$,} \\
1 & \mbox{if $i \neq j$.} \end{array} \right . \]
By $D_{n}$, we mean the $n \times n$ matrix $D_{(n, n)}$.
\item The {\it discriminant} $\Delta  (z) \in {\Bbb Q}[z_{1}, \dots, z_{N}]$ is
defined to be $\delta (z, 2 D_{N})$, {\it i.e.},
\[ \Delta (z)\, := \, \prod_{1 \leq i < j \leq N} (z_{i} - z_{j})^2 .\]
\item Let $m$ be a positive integer and let $\sigma \in S_{m}$ denote
the $m$-cycle $(1 2 \cdots m)$. Given an ordered $m$-tuple
\[ {\frak a} \, := \, (a (1), \dots , a (m)), \] let $\mbox{cirmat} ({\frak a}) $
denote the $m \times m$ {\it circulant} matrix $[c_{ij} ]$ determined by
${\frak a}$, {\it i.e.}, for $1 \leq i, j \leq m$, let
\[ c_{ij} \, := \,a (\sigma ^{1 - i} (j)) . \]
\item Let $m$, $n$ be positive integers such that $m n = N$. Let $a$, $c$ be indeterminates.
Let ${\frak u} := (u (1), \dots , u (m))$ be defined by
\[u (i) \, := \, \left \{ \begin{array}{ll} 2 c & \mbox{if $1 \leq i \leq \frac{m-1}{2}$,} \\
0 & \mbox{otherwise.} \end{array} \right . \]
Let $M_{0} (m, n, a, c)$ be the $N \times N$ symmetric matrix defined as an
$n \times n$ block-matrix $[M_{ij}]$, where, for $1 \leq i, j\leq n$,
\[ M_{ij} \, := \, \left \{\begin{array}{ll} 2 a D_{m} & \mbox{if $i =j$,} \\
\mbox{cirmat} ({\frak u}) & \mbox{if $i < j$,} \\
\mbox{cirmat} ({\frak u})^{T} & \mbox{if $i > j$.} \end{array} \right . \]
\end{enumerate}

\noindent \underline {\bf Examples}:
\[ M_{0} (3, 2, a, c) \, = \, \mat{6}{0 & 2 a & 2 a & 2 c & 0 & 0 \\
2 a & 0 & 2 a & 0 &  2 c & 0 \\ 2 a & 2 a & 0 & 0 & 0 & 2 c \\
2 c & 0 & 0 & 0 & 2 a & 2 a \\ 0 & 2 c & 0 & 2 a & 0 & 2 a \\
0 & 0 & 2 c & 2 a & 2 a & 0}  \]
and  \[ M_{0} (5, 2, a, c) \,
= \, \mat{2}{ 2 a D_{5} & U  \\ U^{T} & 2 a D_{5}} , \]
where
\[ U  \, := \, \mat{5}{2 c & 2 c & 0 & 0 & 0 \\ 0 & 2 c & 2 c & 0 & 0 \\
0 & 0 & 2 c & 2 c & 0 \\ 0 & 0 & 0 & 2 c & 2 c  \\ 2 c & 0 & 0 & 0 & 2 c} . \] \\

We shall need the following preliminary result from \cite{MQS} whose proof we include for completeness.\\

\noindent \underline {\bf Theorem 2}: Let $m$, $n$, $N$ be integers such that
$2 \leq m \leq n \leq N$. Let $k$ be
a field containing ${\Bbb Q}$ and let $z_{1}, \dots, z_{N}$ be indeterminates.
\begin{description}
\item[(i)] Let $n$ be a positive integer and for $1 \leq i \leq n$, let
$g _{i} \in {\Bbb Q}(z_{1}, \dots , z_{N})$ be such that $g_{1} \neq 0$.
Then $g_{1}^2 + g_{2}^2 + \cdots + g_{n}^2 \neq 0$. In particular, given a
$ 0 \neq g \in {\Bbb Q}(z_{1}, \dots , z_{N})$ and a nonempty subset
$S \subseteq S_{N}$, we have
\[ \sum _{\sigma \in S} g (z_{\sigma (1)}, \dots , z_{\sigma (N)})^{2} \, \neq  \, 0 . \]
\item[(ii)] Let $m, n, a, c$ be positive integers such that
$3 \leq m \leq n m = N$ and $m$ is odd. Then, letting $M_{0} := M_{0}(m, n, a, c)$,
 we have
\[ M_{0} \in E (N, (2 a + c n - c)(m - 1)) \] and
\[ Symm_{N} ( \delta (z, M_{0}) ) \,  \neq  \, 0 . \]
\end{description}

\noindent \underline{\bf Proof}: To prove (i), let $h := g_{1}^2 + g_{2}^2 + \cdots + g_{n}^2$.
For $1 \leq  i \leq n$, let $p_{i} , q_{i} \in {\Bbb Q}[z_{1}, \dots , z_{N}]$
be polynomials such that $g_{i} q_{i} = p_{i}$ and $q_{i} \neq 0$. Note that
$g_{1} \neq 0$ implies $p_{1} \neq 0$. Now since $f:= p_{1} q_{1} q_{2} \cdots q_{n}$ is a nonzero
polynomial, there exists $(a_{1}, \dots , a_{N}) \in {\Bbb Q}^{N}$ such that
$f(a_{1}, \dots , a_{N}) \neq 0$. Fix such $(a_{1}, \dots , a_{N})$ and let
$c_{i} := g_{i} (a_{1}, \dots , a_{N})$ for $1 \leq i \leq n$. Then each $c_{i}$ is a
rational number and $c_{1} \neq 0$. Since $c_{1}^2 > 0$ and $(c_{2}^2 + \cdots + c_{n}^2) \geq 0$,
we have $h (a_{1}, \dots , a_{N}) > 0$. In particular, $h  \neq 0$. This proves (i).

To prove (ii), let $M_{ij}$ denote the $ij$-th $m \times m$ block of $M_{0}$ (as in the definition
of $M_{0}$). If $1 \leq i < j \leq n$, then $M_{ij}$ being a circulant matrix and $m$ being odd, each
row-sum as well as each column-sum of $M_{ij}$ is exactly $ c (m - 1)$. Now it is easily verified that
$M_{0}$ is a member of $E (N, (2 a + c n - c)(m - 1))$. Since each entry of $M_{0}$ is a nonnegative
even integer, there exists a nonzero polynomial $g  \in k[z_{1}, \dots , z_{N}]$ such that
\[ Symm _{N} (\delta (z, M_{0})) \,  =  \, \sum _{\sigma  \in S_{N}} \sigma (g (z_{1}, \dots , z_{N}))^2 . \]
Therefore, (ii) follows from (i). \hfill \qed \\

\noindent \underline {\bf Definitions}: Let $m$, $n$ be positive integers and
let $A := [a(i,j)]$ be an $m \times n$ matrix with nonnegative integer entries.
Let $T_{1}, \dots , T_{m}$ be indeterminates and let $T$ stand for $(T_{1}, \dots , T_{m})$.
\begin{enumerate}
\item By $\mbox{max}(A)$, we mean
$\mbox{max} \{a(i,j) \, \mid\, 1 \leq i \leq m, \;\; 1\leq j \leq n \}$.
\item For $1 \leq r \leq m$, define
\[ co(r, A) \, := \, \{j\,\mid\, 1 \leq j \leq n \;\;\mbox{and}\;\; a(r, j) = \mbox{max}(A) \}\]
and let $|co(r, A)|$ denote the cardinality of $co(r, A)$. Let
\[ co(A) \, := \, \bigcup _{r = 1}^{m} co(r, A) . \]
\item For $1 \leq r \leq m$, define
\[ sp(r, A) \, := \, \{j\,\mid\, 1 \leq j \leq n \;\;\mbox{and}\;\; a(r, j) \neq 0 \} \]
and let $|sp(r, A)|$ denote the cardinality of $sp(r, A)$. Let
\[ sp(A) \, := \, \bigcup _{r = 1}^{m} sp(r, A) . \]
\item For $1 \leq r \leq m$, define
\[b (r, A) \, := \, \sum_{j \in sp(r, A) \setminus co(A)} a(r, j) .\]
\item For $1 \leq r < s \leq m$, define
\[\nu (r, s, A) \, := \, \sum_{j \in co(s, A)} a(r, j) + \sum_{j \in co(r, A)}  a(s, j).\]
\item Define
\[ pol(A, T)\, := \, \prod _{r = 1}^{m} T_{r}^{b (r, A)}
 \prod _{1 \leq r < s \leq m} (T_{r} - T_{s})^{\nu(r, s, A)}.\]
\item As usual, let $S_{m}$ denote the permutation group of $\{1, \dots , m\}$. Given a polynomial
$f(T_{1}, \dots, T_{m}) \in {\Bbb Q}[T_{1}, \dots , T_{m}]$ and a permutation $\theta \in S_{m}$,
by $\theta (f (T))$, we mean the polynomial $f(T_{\theta (1)}, \dots , T_{\theta (m)})$. Define
\[grp(A) \, := \, \{ \theta \in S_{m} \, \mid \, |co(r, A)| = |co (\theta (r), A)| \;\;\mbox{for $1 \leq r \leq m$} \} \]
and set
\[ rat(A, T) \, := \, \{ \theta (pol(A, T))^{-1}  \, \mid \, \theta \in grp(A)  \} . \]
\item For an $r \times s$ matrix $A := [a_{ij}]$, define the {\it norm} of $A$ to be
\[ \| A \| \, := \, \sum_{j = 1}^{s} \sum _{i = 1}^{r} a_{ij} . \]
\item We say $A$ is {\it admissible} provided the following three requirements are satisfied.
\begin{description}
\item[(1)] $co(r, A) \neq \emptyset$ for $1 \leq r \leq m$ and
\[ co(r, A) \cap co(s, A)  \, =  \, \emptyset \;\;\;\mbox{for $1 \leq r < s \leq m$.} \]
\item[(2)] $rat(A, T)$ is ${\Bbb Q}$-linearly independent.
\item[(3)] If $M$ is a $p \times q$ submatrix of $A$ with $p, q \geq 2$ and $p + q - 1 = |co (r, A)|$
for some $r$, then $\| M \| < (p + q - 1) \mbox{max}(A)$. \\
\end{description}
\item Given a subset $B$ of $\{1, 2, \dots , N \}$, let
\[ \pi (B) \, := \, \{ (i, j) \in B \times B \, \mid \, i < j \}.  \]
The set $\pi (B) $ is tacitly identified with the set of all $2$-element
subsets of the set $B$, {\it i.e.},
\[ \pi (B) \, = \, \{ \{i, j \} \, \mid \, i, j \in B\;\;\mbox{and $i \neq j$} \} .\]
By $\pi [N]$, we mean the set $\pi (\{1, \dots , N \})$.
\item Given a subset $C \subseteq \pi [N]$ and a function
$\varepsilon : C \rightarrow {\Bbb N}$, the image of $(i, j) \in C$
via $\varepsilon $ is denoted by $\varepsilon (i, j)$.
A nonnegative integer $w$ is identified with the constant
function $C \rightarrow {\Bbb N}$ that maps each member of $C$ to $w$.
\end{enumerate}

\noindent \underline {\bf Remarks}:
\begin{enumerate}
\item Let $A$ be an $m \times n$ matrix with nonnegative integer entries. If $m = 1$, then $A$ is admissible.
If $m  \geq n + 1$, then $A$ is not admissible. Even when $m =n$, admissibility of $A$ need
not guarantee admissibility of $A^{T}$.
\item Let $A$ be an $m \times n$ matrix with nonnegative integer entries satisfying requirements (1)
and (2) in the definition of admissibility. If $|co (r, A)| \leq 2$ for $1 \leq r \leq m$, then
$A$ is admissible.
\item Let $A$ be an $m \times n$ matrix with nonnegative integer entries satisfying requirements (1)
and (3) in the definition of admissibility. If $|co(r, A)| \neq |co(s, A)|$ for $1 \leq r < s \leq m$,
then $A$ is easily verified to be admissible. \\
\end{enumerate}

\noindent \underline {\bf Examples}:
\begin{enumerate}
\item Let
\[ A \, := \, \mat{5}{0 & 2 & 1 & 0 & 2 \\ 2 & 1 & 0 & 2 & 1 \\ 1 & 0 & 2 & 1 & 0}. \]
Then, $co(1, A) = \{2, 5\}$, $co(2, A) = \{1, 4\}$ and $co(3, A) = \{3\}$. Hence
$grp(A) = \{ id, \, \tau \}$, where $id$ is the identity permutation and $\tau$ denotes
the transposition $(1,2)$. Also, we have
\[pol(A, T) = (T_{1} - T_{2})^{2}(T_{1} - T_{3}) (T_{2} - T_{3})^{2}. \]
It is straightforward to verify that $rat(A, T)$ is the set
\[ \left \{ \frac{1}{(T_{1} - T_{2})^{2}(T_{1} - T_{3}) (T_{2} - T_{3})^{2}}, \;\;\;\;
\frac{1}{(T_{2} - T_{1})^{2}(T_{2} - T_{3}) (T_{1} - T_{3})^{2}}  \right \},  \]
which is ${\Bbb Q}$-linearly independent. So, $A$ is admissible.
\item We leave it to the reader to verify that if
\[A \, := \, \mat{3}{2 & 1 & 1 \\ 0 & 2 & 1 \\ 0 & 0 & 2}, \]
then $grp(A) = S_{3}$, $pol(A, T) = (T_{1} - T_{2})(T_{1} - T_{3}) (T_{2} - T_{3})$,
and
 \[ rat(A, T) \, = \, \left \{\frac{1}{pol(A, T)}, \; \;\; \frac{-1}{pol(A, T)} \right \} \]
is ${\Bbb Q}$-linearly dependent. So, $A$ is {\em not} admissible. \\
\end{enumerate}

\noindent \underline {\bf Theorem 3}: Let $m$, $N$ be integers such that
$1 \leq m \leq N - 2$. As before, $k$ is a field containing ${\Bbb Q}$,
$z_{1}, \dots, z_{N}$ are indeterminates and $z$ stands for $(z_{1}, \dots , z_{N})$.
Let $A := [a(i, j)]$ be an $m \times (N - m )$ matrix with nonnegative integer
entries $a(i, j)$. Assume that $\mbox{max}(A) =  2 a$ for some positive integer $a$ and $A$
is admissible. Let $E \in E(N)$ be the matrix given in block-form by
\[ E \, := \, \mat{2}{0 & A \\ A^{T} & 0}. \]
Then, we have
\[ Symm_{N} \left ( \frac{1}{\delta(z, E)} \right ) \, \neq \, 0. \] \\

\noindent \underline{\bf Proof}: Let
$\mu (z) := \delta (z, E)$ and denote the $(i, j)$-th entry of $E$ by  $\varepsilon (i, j)$. Then
\[\mu (z) \, := \, \prod _{(i, j) \in \pi [N]} (z_{i} - z_{j})^{\varepsilon (i, j)} .\]
Let $J := J_{1} \cup \cdots \cup J_{m}$ and $B := B_{1} \cup \cdots \cup B_{m}$, where
\[ J_{r} \, := \, \{r \} \cup \{ m + j \, \mid \, j \in sp(r, A) \} \;\;\; \mbox{and} \;\;\;
 B_{r} \, :=  \,  \{r \} \cup \{ m + j \, \mid \, j \in co(r, A) \} \]
for $1 \leq r \leq m$. Then $B_{r} \cap B_{s} \, = \, \emptyset$ for $1 \leq r < s \leq m$.
Define \[ G \,:= \, \{ \sigma \in S_{N}  \, \mid \, \sigma (J) = J \} \] and let $H \subseteq G$ be the set
of all $\sigma \in G$ such that for each $B_{r}$, there is an $s$ (depending on $r$ and $\sigma$ but
necessarily unique) with $\sigma(B_{r}) \subseteq B_{s}$. Note that $H$ contains at least the identity
permutation. Moreover, it is straightforward to verify that a $\sigma \in H$ determines a unique
permutation $\theta$ of $\{1, \dots , m\}$ such that $\sigma (B_{i}) = B_{\theta (i)}$ for $1 \leq i \leq m$
and then clearly we have $\theta \in grp(A)$. The permutation $\theta$ is said to be {\it induced} by $\sigma$.

Next, let $t$, $y$, $T_{1}, \dots , T_{m}$ and $x_{1}, \dots , x_{N}$ be indeterminates. Let $x$ stand for
$(x_{1}, \dots , x_{N})$ and $T$ for $(T_{1}, \dots , T_{m})$. Let
\[ \alpha : k [z] \rightarrow k [t, y, T, x, z] \]
denote the $k$-homomorphism of rings defined by
\[ \alpha (z_{i}) \, := \, \left \{ \begin{array}{ll} t x_{i} + T_{r}  &
\mbox{if $i \in B_{r}$ with $1 \leq r \leq m$,} \\
y x_{i}  &  \mbox{if $i \in J \setminus B$,} \\
 z_{i} &  \mbox{otherwise,} \end{array} \right .   \]
for $1 \leq i \leq N$. Since $B_{1}, \dots, B_{m}$ are pairwise disjoint, $\alpha$ is well-defined.
Also, $\alpha$ is easily seen to be injective and hence naturally extends to an injective
field homomorphism $k (z) \rightarrow  k (t, y, T, x , z)$, which we will also
denote by $\alpha$.

For $\sigma \in G$, $1 \leq i \leq m$ and $j \in J \setminus \{1, \dots , m \}$, we have
\[ \alpha (z_{\sigma (i)} - z_{\sigma (j)}) \, = \, \left \{ \begin{array}{ll}
t (x_{\sigma(i)} - x_{\sigma (j)}) + T_{r} - T_{s} &
\mbox{if $\sigma (i) \in B_{r}$, $\sigma (j) \in B_{s}$, } \\
t x_{\sigma(i)} + T_{r} - y x_{\sigma (j)} &
\mbox{if $\sigma (i) \in B_{r}$, $\sigma (j) \in J \setminus B $, } \\
y x_{\sigma(i)} -  t x_{\sigma (j)} - T_{r} &
\mbox{if $\sigma (i) \in J \setminus B$, $\sigma (j) \in B_{r}$,} \\
y x_{\sigma (i)} - y x_{\sigma (j)} &
\mbox{if $ \sigma (i) \in J \setminus B$, $\sigma (j) \in J \setminus B$.}
\end{array} \right . \]
If $\sigma \in G$, then the total $z$-degree of $\alpha ( \sigma( \mu (z)))$ is $0$.
Given $\sigma \in S_{N} \setminus G$, let $s \in J$ be such that $\sigma (s)$ is not in $J$.
If $1 \leq s \leq m$, then for any $j \in sp (s, A)$, the $z$-degree of
$\alpha ( z_{\sigma (s)} - z_{\sigma (m + j)})$ is $1$. On the other hand, if $m < s$, then
for an $r \in \{1, \dots , m \}$ such that $s \in J_{r}$, the $z$-degree of
$\alpha ( z_{\sigma (r)} - z_{\sigma (s)} )$ is $1$. It follows that the $z$-degree of
$\alpha ( \sigma (\mu (z)))$ is $\geq 1$ if and only if $\sigma \in S_{N} \setminus G$.

For $1 \leq r \leq m$, define
\[\lambda (r) \, := \,\prod _{j \in co (r, A)} (z_{r} - z_{m + j})\;\;\;\mbox{and}\;\;\;
\psi (r) \, := \, \prod _{j \in J_{r} \setminus B} (z_{r} - z_{j})^{a (r, j)} .\]
For $1 \leq r < s \leq m$, define
\[\phi (r, s) \, := \,
\prod _{j \in co (s, A)} (z_{r} - z_{m + j})^{a (r, j)} \prod _{j \in co (r, A)} (z_{s} - z_{m + j})^{a (s, j)} .\]
Clearly, we have $\mu (z) \in {\Bbb Q} [z]$ and
\[ \mu (z) \, = \, \left ( \prod _{r=1}^{m} \lambda (r) ^{2 a}  \psi (r) \right )
\prod _{1 \leq r < s \leq m} \phi (r, s). \]
Define $g \in {\Bbb Q} [x]$ by setting
\[g(x_{1}, \dots , x_{N}) \, := \,\prod _{r=1}^{m} \prod _{j \in co (r, A)} \, (x_{r} - x_{m + j}) \]
and for $\sigma \in S_{N}$, let $\sigma (g(x))$ denote the polynomial $g(x_{\sigma(1)}, \dots , x_{\sigma (N)})$.
For $\sigma \in G$, let $Q_{\sigma} (t, y, T, x) := \alpha ( \sigma (\mu (z)))$ and
\[ P_{\sigma} (t, y, T, x) \, := \, \prod _{r=1}^{m} \alpha (\sigma (\psi (r)))
 \prod _{1 \leq r < s \leq m} \alpha ( \sigma (\phi (r, s))). \]
Then, we have
\[Q_{\sigma} (t, y, T, x) \, = \, P_{\sigma} (t, y, T, x) \prod _{r=1}^{m} \alpha (\sigma (\lambda (r)))^{2 a}.\]

Let $id$ denote the identity permutation and $b$ the cardinality of $co(A)$. Observe that
\[Q_{id} (t, y, T, x) \, = \, t ^{2 a b} \,g(x)^{2 a}\, P_{id} (t, y, T, x) \]
and $P_{id} (0, 0, T, x) = c \cdot pol(A, T)$, where $c \in \{ -1, 1 \}$. Moreover, if $\sigma \in H$ and
$\theta \in grp(A)$ denotes the permutation induced by $\sigma$, then
\[Q_{\sigma} (t, y, T, x) \, = \, t ^{2 a b}\, \sigma (g(x))^{2 a} \,P_{\sigma} (t, y, T, x) \]
and $P_{\sigma} (0, 0, T, x) \, := \, c \cdot \theta ( pol(A, T))$. Here, it is important to notice that
$c$ is a nonzero integer that does not depend on the choice of $\sigma$. Now letting
\[\alpha(h) \, := \, \sum _{\sigma \in H} \frac{1}{ \alpha( \sigma( \mu (z)))} \, = \,
\sum _{\sigma \in H} \frac{1}{Q_{\sigma} (t, y, T, x)}, \]
we can substitute $t = y = 0$ in the resulting expression of $t^{2 a b} \alpha (h)$ to get
\[ (*)\;\;\;\;\;c \cdot \sum _{f \in \, rat(A, T)} \left ( \sum _{*} \frac{1}{\sigma (g(x))^{2 a}} \right ) f ,\]
where the inner sum (in $*$) is over all $\sigma \in H$ such that their induced $\theta$ in $grp(A)$
satisfies $ \theta (pol(A, T)) = 1 / f$. By (i) of Theorem 2, this inner sum is nonzero provided
it is nonempty. In particular, the inner sum is nonzero when $f = 1 / pol (A, T)$. Since $A$ is admissible,
$rat(A, T)$ is a ${\Bbb Q}$-linearly independent subset of ${\Bbb Q}(T)$ and hence, as a subset of ${\Bbb Q}(T,x)$,
$rat(A, T)$ is linearly independent over the field ${\Bbb Q}(x)$. Thus, the above sum $(*)$ is nonzero. It follows that
$t^{2 a b} \alpha (h)$ has $t$-order $0$. Consequently, $\alpha (h)$ has $t$-order $- 2 a b$ (a negative even integer).

Next, fix a $\sigma \in G \setminus H$. Firstly, we have
\[\sigma (\mu (z)) \, = \, \pm \prod _{(i, j) \in \pi [N]}
(z_{i} - z_{j})^{\varepsilon ( \sigma^{-1} (i), \, \sigma^{-1} (j))} .\]
Secondly, given $(i, j) \in \pi [N]$, it is evident that $\alpha (z_{i} - z_{j})$ has positive
$t$-order if and only if $(i, j)$ is in $\pi (B_{r})$ for some $r$. Hence, the $t$-order of
$\alpha (\sigma (\mu (z)))$ equals
\[ \sum _{r = 1}^{m} \sum _{(i, j) \in \pi (B_{r})} \varepsilon ( \sigma^{-1} (i), \, \sigma^{-1} (j)) \; = \;
\sum _{r = 1}^{m} \sum _{(i, j) \in \pi (\sigma ^{-1} (B_{r}))} \varepsilon (i, \, j) . \]
For $1 \leq r \leq m$, define
\[ R_{r} (\sigma) \, :=  \,\{1 , \dots , m \} \cap \sigma^{-1} (B_{r}) \;\;\;\mbox{and}\;\;\;
K_{r}(\sigma) \, := \, \sigma^{-1} (B_{r}) \setminus \{1, \dots , m \} . \]
Clearly, the sets $R_{r} (\sigma)$, $K_{r} (\sigma)$ partition $\sigma ^{-1} (B_{r})$; in particular,
the sum of their cardinalities equals $1 + |co (r, A)|$. Given $(i, j) \in \pi [N]$, we have
\[ \varepsilon (i, j) \,  = \, \left \{\begin{array}{ll} a(i, j - m) &
\mbox{if $1 \leq i \leq m$ and $m + 1 \leq j \leq N$,} \\ 0 & \mbox{otherwise.} \end{array} \right . \]
Hence, for $1 \leq r \leq m$, letting $M_{r} (\sigma)$ denote the submatrix of $A$ determined by row-set
$R_{r} (\sigma)$ and column-set $K_{r} (\sigma)$, we have
\[\sum _{(i, j) \in \pi (\sigma ^{-1} (B_{r}))} \varepsilon (i, \, j) \, = \,
\sum _{(i, j) \in R_{r} (\sigma) \times K_{r} (\sigma)} a(i, \, j - m) \, = \, \| M_{r} (\sigma) \|. \]
If an $M_{r} (\sigma)$ is empty, then $\| M_{r} (\sigma) \| = 0$.
If an $M_{r} (\sigma)$ has only one row, then clearly $\| M_{r} (\sigma) \| \leq 2 a | co(r, A) |$.
If an $M_{r} (\sigma)$ has a single column and at least two rows, then the admissibility-condition (1)
implies $\| M_{r} (\sigma) \| <  2 a | co(r, A)|$. If an $M_{r} (\sigma)$ has two or more rows as well as
two or more columns, then the admissibility-condition (3) guarantees that
$\| M_{r} (\sigma) \| < 2 a | co(r, A) |$. These observations allow us to infer that
\[ (**) \;\;\;\;\;\;\;\; \;\sum _{r = 1}^{m} \| M_{r} (\sigma) \| \, \leq \, 2 a b \]
and, in view of the admissibility-condition (1) satisfied by $A$, that equality holds only when for $1 \leq r \leq m$,
$M_{r} (\sigma)$  is a row-matrix with each entry equal to $2 a$, {\it i.e.}, only when there is a  permutation $\theta$
of $\{1, \dots , m \}$ such that for $1 \leq r \leq m$,
\[ R_{r} (\sigma)  \, =  \, \{ \theta (r) \} \;\;\;\;\mbox{and} \;\;\;\; \sigma ^{-1} (B_{r})\, \subseteq\, B_{\theta (r)}. \]
Since $\sigma$ is not in $H$, its inverse is not in $H$ and hence $(**)$ must be a strict inequality. We have thus
proved that for each $\sigma \in G \setminus H$, the polynomial $\alpha (\sigma (\mu (z)))$ has $t$-order strictly less
than $2 a b$.

Now, define
\[\gamma := \sum _{\sigma \in G} \frac{1}{\sigma(\mu (z))}, \]
and note that
\[ \alpha (\gamma) \, = \, \alpha (h) + \sum _{\sigma \in G \setminus H} \frac{1}{\alpha (\sigma (\mu (z)))}. \]
Since for each $\sigma \in G \setminus H$, the $t$-order of $\alpha ( \sigma (\mu (z)))$
is strictly less than $2 a b$,  the $t$-order of $\alpha (\gamma)$ is the negative integer $ - 2 a b$.
Thus, $\alpha (\gamma)$ is a nonzero element of the field $k(t, y, T, x)$. If $J = \{1, \dots , N \}$, then $G = S_{N}$
and since $\gamma \neq 0$, we have established the desired result. Henceforth, assume that $J \neq \{1, \dots , N \}$.
Apply $\alpha$ to both sides of the equation
\[ Symm_{N} \left (\frac{1}{\mu (z)} \right )\, = \, \gamma +
\sum _{\sigma \in S_{N} \setminus G} \frac{1}{\sigma(\mu (z))}. \]
Let $d(\sigma)$ for $\sigma \in S_{N} \setminus G$ denote the total $z$-degree of
$\alpha ( \sigma( \mu (z)))$. As observed earlier, $d (\sigma) \geq 1$ for $\sigma \in S_{N} \setminus G$.
Let $d$ be the sum of $d(\sigma)$ as $\sigma$ ranges over $S_{N} \setminus G$. Define
\[ \eta \, :=  \, \prod_{\sigma \in S_{N} \setminus G} \alpha ( \sigma(\mu (z))). \]
Since $\alpha (\gamma)$ is in $k (t, y, T, x)$, the product $\alpha (\gamma) \eta$ has
total $z$-degree exactly $d$. On the other hand,
\[ \sum _{\sigma \in S_{N} \setminus G} \frac{\eta }{\alpha ( \sigma( \mu (z)))} \]
is a polynomial whose total $z$-degree is at most $d - 1$. Consequently,
\[ Symm_{N} \left (\frac{1}{\mu (z)} \right )\, \neq \, 0 .\]
Our assertion is thus fully established. \hfill \qed \\

\noindent \underline {\bf Corollary}: Let $m$, $N$ be as in the above theorem and let
$A := [a(i, j)]$ be an $m \times (N - m )$ matrix with nonnegative integer entries $a(i, j)$.
\begin{enumerate}
\item Assume that the following holds.
\begin{description}
\item[(i)] There is a positive integer $a$ such that $\mbox{max}(A) = 2 a$ and
$co(r, A) \neq \emptyset$ for $1 \leq r \leq m$.
\item[(ii)] For $1 \leq j \leq N - m$, we have
\[ \sum_{i=1}^{m} a(i, j) \, \leq \, 2 a, \]
{\it i.e.}, each column-sum of $A$ is at most $ 2 a$.
\end{description}
Then, letting $E \in E(N)$ be defined as in the above theorem, we have
 \[ Symm_{N} ( \delta (z, -E))\,\neq \, 0.  \]
\item Assume that the following holds.
 \begin{description}
\item[(i)] There is a positive integer $a$ such that $\mbox{max}(A) = 2 a$.
\item[(ii)] $|co(r, A)| = 1$ for $1 \leq r \leq m$ and
\[ co(r, A) \cap co(s, A)  \, =  \, \emptyset \;\;\;\mbox{for $1 \leq r < s \leq m$.} \]
\item[(iii)] There is a nonnegative integer $b < 2 a$ such that for $1 \leq i, \,r \leq m$ with
$i \neq r$ and $j \in co (r, A)$, we have $a(i, j) = b $.
\end{description}
Then, letting $E \in E(N)$ be defined as in the above theorem, we have
 \[ Symm_{N} ( \delta (z, -E))\,\neq \, 0.  \]
\item Let $a$, $b$, $c$, $r$, $s$ be positive integers such that $b < 2 a \leq 2 c$ and
$r \leq s \leq N - 1$. Suppose $A := [u_{1}, \dots , u_{N - 1}]$ is the $1 \times N - 1$
matrix such that $u_{i} := 2 a$ for $1 \leq i \leq r$, $u_{i} := b$ for $r + 1 \leq i \leq s$, and
$u_{i} = 0$ for $s + 1 \leq i \leq N - 1$. Let $E \in E(N)$ be defined as in the above theorem.
Then, letting $E_{(r, s)} (N; a, b, c) \, := \, 2 c D_{N}  - E$, we have
\[ Symm_{N} ( \delta (z, E_{(r, s)} (N; a, b, c))) \,\neq \, 0. \]

\item Assume that the following holds.
\begin{description}
\item[(i)] $N = 2 m$ and $m \geq 2$.
\item[(ii)] $a(i, j) = a(j, i)$ for $1 \leq i < j \leq m$.
\item[(iii)] There are positive integers $a$, $a_{1}, \dots , a_{m}$ such that
\[ a(i, i)\, = \, 2 a_{i} \, \geq  \, 2 a \, > \, a(i, j)\;\;\;\;\mbox{for $1 \leq i < j \leq m$.} \]
\end{description}
Then, letting $E \in E(N)$ be defined as in the above theorem, we have
 \[ Symm_{N} ( \delta (z, -E))\,\neq \, 0.  \]
\end{enumerate}

\noindent \underline{\bf Proof}: To prove the first two assertions, it suffices to show that under their
respective hypotheses, $A$ is admissible.

Suppose $A$ satisfies the requirements of 1. Now hypothesis (ii) of 1 ensures that if $j \in co(A)$, then there
is only one nonzero entry in the $j$-th column of $A$ and that nonzero entry is $2 a$. Thus, we have
$\nu (r, s, A) = 0$ for $1 \leq r < s \leq m$. It follows
that \[ rat(A, T) \,  =  \,
\left \{ T_{\theta (1)}^{-b (1, A)} \cdots \cdots T_{\theta (m)}^{-b (m ,A)} \, \mid \, \theta \in grp (A) \right \} \]
and hence $rat (A, T)$ is a ${\Bbb Q}$-linearly independent subset of ${\Bbb Q} (T)$. Next, let $M$ be a
$p \times q$ submatrix of $A$, where $p \geq 2$. Since hypothesis (iii) of 1 ensures that $\| M \| \leq 2 a q$ and
$ 2 a q < 2 a (q + 1) \leq 2 a (p + q - 1)$, our matrix $A$ is indeed admissible.

Secondly, assume that $A$ satisfies the requirements of 2. Then hypothesis (iii) of 2 ensures that
$\nu (r, s, A) = 2 b$ for $1 \leq r < s \leq m$. Consequently,
\[ rat(A, T) \,  =  \,
\left \{ \Delta (T) ^{-b} \cdot T_{\theta (1)}^{-b (1, A)} \cdots \cdots T_{\theta (m)}^{-b (m ,A)}
\, \mid \, \theta \in grp (A) \right \}. \]
It is straightforward to verify that $rat (A, T)$ is a ${\Bbb Q}$-linearly independent subset of
${\Bbb Q} (T)$ and $A$ is admissible.

Assertion 3 follows from the fact that
$\delta(z, 2 c D_{n}) = \Delta (z) ^{c}$ is symmetric and
\[ Symm_{N}(\delta (z, 2 c D_{n} - E) \, = \, \Delta  (z)^{c} \cdot Symm_{N}(\delta (z, - E)), \]
with $Symm_{N}(\delta (z, - E))$ nonzero by assertion 1.

The remainder of the proof establishes assertion 4. Letting $\mu (z) := \delta (z, E)$, in view of
our hypothesis (ii), we have
\[\mu (z) \, := \, \prod _{1 \leq i \leq m } (z_{i} - z_{m + i})^{2 a_{i}}
\prod _{1 \leq i  <  j  \leq m} [(z_{i} - z_{m + j}) (z_{j} - z_{m + i})]^{a(i, j)} .\]
Let $J := \{(i, j) \, \mid \, 1 \leq i \leq m < j \leq N \}$. For $\sigma \in S_{N}$, define
$B_{\sigma}$ to be the set of $(i, j) \in \pi [N]$ such that $\sigma (i, j) = (r, m + r)$ for
some $1 \leq r \leq m$. Let $B := B_{\iota} = \{(r, m + r)\, \mid \, 1 \leq r \leq m \}$ and
\[ G \, := \, \{ \sigma \in S_{N} \, \mid \, B_{\sigma} = B \}. \]
Note that $B \subset J$ and the identity permutation belongs to $G$.

Let $x$ and $T$ be as before and let
\[ \alpha : k [z] \rightarrow k [t, T, x] \]
denote the $k$-homomorphism of rings defined by
\[ \alpha (z_{i}) \, := \, \left \{ \begin{array}{ll} t x_{i} + T_{i}  &
\mbox{if $1 \leq i \leq m$,} \\
t x_{i} + T_{i - m}  &  \mbox{if $m + 1 \leq i \leq N$.} \end{array} \right .  \]
Then $\alpha$ is easily seen to be injective and hence it naturally extends to an injective
field homomorphism $k (z) \rightarrow  k (t, T, x)$, which (by a slight abuse of notation) is also
denoted by $\alpha$.

For $1 \leq i, j \leq m$, let $q_{1}, q_{2} \in k[z]$ be the polynomials
\[ q_{1}(i, j) \, := \,(z_{i} - z_{m + j}) (z_{j} - z_{m + i}), \;\;\;
q_{2}(i, j) \, := \,(z_{i} - z_{j}) (z_{m + j} - z_{m + i}). \]
Note that $q_{1} (i, j) = q_{1} (j, i)$, $q_{2} (i, j) = q_{2} (j, i)$, and
$q_{2} (i, i) = 0$. Evidently,
\[ \mu (z) \, = \, \prod _{1 \leq i  \leq m } q_{1}(i, i)^{a_{i}}
\prod _{1 \leq i < j \leq m } q_{1}(i, j)^{a(i, j)} .\]
Fix $\sigma \in G$ and $(i, j) \in \pi [m]$. Clearly, $\sigma (p, m + p) \in B$ for all
$(p, m + p) \in B$ and hence $\{\sigma (i), \sigma (m + i) \} = \{r, m + r \}$ for some
$1 \leq r \leq m$. Likewise, $\{\sigma (j), \sigma (m + j) \} = \{s, m + s \}$ for some
$1 \leq s \leq m$. Since $i \neq j$, we have $\{r, m + r \} \cap \{s, m + s \} = \emptyset$.
Clearly, $\alpha (\sigma (q_{1} (i, i))) = t^{2} (x_{i} - x_{m+i})^2$. Observe that
$\sigma (q_{1}(i, j)) \in \{ q_{1}(r, s), \, q_{2}(r, s) \}$ and hence if $i \neq j$, then
\[ \alpha (\sigma (q_{p} (i, j))) \, = \, (T_{r} - T_{s})^{2} \;\mbox{mod}\; t  \;\;\;\;
\mbox{for $1 \leq p \leq 2$.} \]

Let $d := 2 (a_{1} + \cdots + a_{m})$. Consider $\sigma \in S_{N} \setminus G$. Then
$| B_{\sigma} \cap B | \leq m - 1$ and $|B_{\sigma} \cap J| \leq m$. From our hypothesis
(iii), it follows that the $t$-order of $\alpha (\sigma (\mu (z)))$ is strictly less than $d$.
On the other hand, if we consider a $\sigma \in G$, then there are polynomials
$P_{\sigma} (x) \in k[x]$ and $Q_{\sigma} (t, x, T) \in k[t, x, T]$ such that
\[ \alpha (\sigma (\mu (z))) \, = \, t^{d} P_{\sigma}(x)^{2} Q_{\sigma} (t, x, T). \]
Moreover, from what was observed above, there is a $h_{\sigma} (T) \in k[T]$ such that
$Q_{\sigma} (0, x, T) = h_{\sigma} (T)^2$. Let
\[v (t, x, T) \, := \, \sum _{\sigma  \in S_{N} \setminus G} \frac {1}{\alpha (\sigma(\mu (z)))}
\;\;\;\; \mbox{and} \;\;\;\; w (t, x, T)\, := \,
\sum _{\sigma \in G} \frac {t^{d}}{\alpha (\sigma(\mu (z)))} .\]
Then $\alpha (Symm_{N} (\delta (z, -E))) = v (t, x, T) + t^{-d} w(t, x, T)$. First, note that the
$t$-order of $v(t, x, T)$ is strictly greater than $ - d$. Secondly, since
\[ w (0, x, T)\,  = \,
\sum _{\sigma \in G} \left ( \frac{1}{P_{\sigma} (x) h_{\sigma} (T)} \right ) ^{2}, \]
assertion (i) of Theorem 2 assures that $w (0, x, T) \neq 0$ and hence $w (t, x, T) \neq 0$.
Consequently, the $t$-order of $v (t, x, T) + t^{-d} w(t, x, T)$ is exactly $-d$. Nonzero-ness
of $Symm_{N} (\delta (z, -E))$ now readily follows. \hfill \qed \\

\noindent \underline {\bf Example}: We present an example which shows that although
assertion 4 of the above Corollary is similar in spirit to Theorem 3, it does
offer something essentially different. Consider the $6 \times 6$ symmetric matrix $A := [a(i, j)]$,
where $a(i, i) = 2$ for $1 \leq i \leq 6$, $a(1, 2) = a(2, 1) = a(3, 4) = a(4, 3) = 0$, and each of
the remaining $a(i, j)$ is $1$. Then $A$ satisfies the admissibility requirements (1) and (3), but a
MAPLE computation shows that $rat (A, T)$ is ${\Bbb Q}$-linearly dependent and thus $A$ is not admissible.
Nevertheless, $A$ does satisfy the hypotheses of assertion 4 of the above Corollary. \\

\section{Applications to configurations of Fermions containing quasielectrons}

We now apply the theorems proved so far to construct the correlation function
$G(z_{1}, \dots , z_{N})$ for a system of $N$ interacting Fermions in several cases.
 Recall that the trial wave function for such a system is
given by the product $F(z_{1}, \dots , z_{N}) G(z_{1}, \dots , z_{N})$, where
\[ F(z_{1}, \dots , z_{N}) \, := \, \prod_{1 \leq i < j \leq N} (z_{i} - z_{j})  \]
is alternating and $G(z_{1}, \dots , z_{N})$ is symmetric in $z_{1}, \dots , z_{N}$. \\

Let $N\geq3$ be an integer and let $m$ be a positive integer not exceeding  $1 + (N / 2)$. Consider a configuration
containing $m$ QEs above the $\nu ={^1/_3}$ IQL state for the rest of the electrons. Given a positive integer $d$, let
\[ {\frak G} (N, d, X) := \frac{ \prod _{i=1}^{d} (1 - X^{N + i})}{\prod _{i=2}^{d}(1 - X^{i})}. \]
Note that ${\frak G} (N, d, X)$ is a polynomial in $X$ of degree $N d + 1$.  Let $q(N, m, X)$ and $r(N, m, X)$ be the unique polynomials
in $\sqrt{X}$ such that
\[{\frak G}(N - 2 m + 2, m, X) \, = \, q(N, m, X) X^{1 + \frac{1}{2} m (N - 2 m + 2)} + r(N, m, X) \]
and the $X$-degree of $r(N, m, X)$ is strictly less than $1 + (m / 2)(N - 2 m + 2)$. Let $\Lambda (N, m)$
denote the support of $q(N, m, X)$, {\it i.e.}, the set of half-integers $\varepsilon$ for
which the coefficient of $X^{\varepsilon}$ in $q(N, m, X)$ is nonzero. Then, from~\cite{BQQW} (or~\cite{CK}),
it follows that $\Lambda (N, m)$ is the set of the possible values of $L$.

Given an $L$ in $\Lambda (N, m)$,
the correlation function $G(z_{1}, \dots , z_{N})$ that we seek to construct is a nonzero homogeneous
polynomial of total degree
\[\kappa _{G} \, := \, \frac{N (2 (N - 1) - m)}{2} - L \]
such that its $z_{i}$-degree is at most  $2 (N - 1) - m$ for $1 \leq i \leq N$. If there are two or more
possible values of $L$, we prefer to denote the corresponding $G$ by $G_{L}$. In order for $0$ to belong to
$\Lambda (N, m)$, it is necessary that $N m$ be even. When $L = 0$, the corresponding
polynomial $G_{0}$ is necessarily a binary invariant of type $(N, 2 (N - 1) - m)$. In contrast, if $L > 0$,
then $G_{L}$ is not a binary invariant; nevertheless, since our $G_{L}$ is
obtained by symmetrizing $\delta (z, E)$ for an $E \in E(N)$, it is indeed a semi-invariant, {\it i.e.},
a homogeneous, symmetric, translation invariant polynomial. In our constructions below, where various $G$
are realized as $Symm_{N}(\delta (z, E))$, we strive to find an $E \in  E(N, \, \leq 2(N - 1) - m)$ having
$2 D_{N - m}$ as a diagonal-block and simultaneously making sure that $\mbox{max}(E)$ is as small as possible.

We will make use of the following further notation. Given $E := [ \varepsilon (i, j) ] \in E(N)$ and an integer $b$, define $frq (b, E)$
({\it frequency of $b$ in $E$}) to be the cardinality of the set
\[ \{(i, j) \, \mid \, 1 \leq i < j \leq N \;\; \mbox{with} \;\; \varepsilon (i, j) = b \}. \]
Given nonnegative integers $d$ and $\lambda$ and positive integers $r$ and $s$, let
${\Bbb M}(r, s, d, \lambda)$ be the set of all $r \times s$ matrices $A := [a_{ij}]$ having nonnegative integer entries
such that $\| A \| = \lambda$,
 \[ \sum _{j = 1}^{s} a_{ij} \, \leq \, d \;\;\; \mbox{for $1 \leq i \leq r$ and}
\;\;\;  \sum _{i = 1}^{r} a_{ij} \, \leq \, d \;\;\; \mbox{for $1 \leq j \leq s$.}  \]\\

We now consider systems of interacting electrons with various numbers of quasielectrons.\\

\begin{enumerate}
\item Suppose first that we have a single QE, {\it i.e.}, $m = 1$. Then
\[ {\frak G}(N - 2 m + 2, m, X) \,  =  \, {\frak G}(N, 1, X) \, = \, (1 - X^{N+1}). \]
Consequently, the only possible value of $L$ in this case is $N / 2$. Let $G:=Symm _{N} (\delta (z, E))$,
where $E := E_{(1, N - 1)} (N; 1, 1, 1)$ (see Corollary of Theorem 3).
Now the third assertion of the Corollary of Theorem 3 ensures that $G(z_{1}, \dots , z_{N})$ is a nonzero
polynomial which is homogeneous of total degree
\[\kappa _{G} \, := \, \frac{N (2 N - 3)}{2} - \frac{N}{2}  \, = \, N (N - 2),\]
and its $z_{i}$-degree is at most $2 N - 3$ for $1 \leq i \leq N$. We have
\[ suppt (E) \, =  \, \{0, 1, 2 \}, \]
\[ frq (b, E) \, = \, \left \{\begin{array}{ll} 1 & \mbox{if $b = 0$,}\\
N - 2 & \mbox{if $b = 1$,} \\
\frac{(N - 1) (N - 2)}{2} & \mbox{if $b = 2$.} \end{array} \right . \]

\item Consider now the case of two QEs, {\it i.e.}, $m = 2$. Then
\[ {\frak G}(N - 2 m + 2, m, X) \,  =  \, {\frak G}(N - 2, 2, X) \, = \,
\frac{(1 - X^{N - 1}) (1 - X^{N})}{(1 - X^{2})}. \]
It is straightforward to verify that
\[\Lambda (N, m) \, = \, \left \{p(N) + 2 r \, \mid \, 0 \leq r \leq \frac{1}{2} (N - 2 - p(N)) \right \}, \]
where
\[ p(N) \, := \, \left \{\begin{array}{ll} 0 & \mbox{if $N$ is even,} \\
1 & \mbox{if $N$ is odd.} \end{array} \right . \]
For $1 \leq i \leq N - 2$, define $R_{i} := [2,\; 0]$ if $i$ is odd and
$R_{i} :=[0, \;  2]$ if $i$ is even. Let $A$ be the $(N - 2) \times 2$ matrix having
$R_{i}$ as its $i$-th row. For $1 \leq r \leq (1/ 2) (N - 2 - p(N))$, let $E_{r} \in E(N)$ be the
matrix defined in block-form by
\[ E_{r} \, := \, \mat{2}{a_{r} D_{2} & A^{T} \\ A & 2 D_{N - 2}},\;\;\;\;
\mbox{where $a_{r} := N - 2 - p(N) - 2 r$.} \]
Assertion (i) of Theorem 2 ensures that for each $L = p(N) + 2 r$, $G_{L} := Symm_{N}(\delta(z, E_{r}))$
is a nonzero polynomial which is homogeneous of total degree
\[\kappa _{G_{L}} \, := \, \frac{N (2 N - 4)}{2} - L \, = \, N (N - 2) - p(N) - 2 r,  \]
and its $z_{i}$-degree is at most $2 N - 4$ for $1 \leq i \leq N$. We have
\[ suppt (E_{r}) \, =  \, \{0, 2, N - 2 - p(N) - 2 r \}, \]
\[ frq (b, E_{r}) \, = \, \left \{\begin{array}{ll} N - 2 & \mbox{if $b = 0$,}\\
\frac{(N - 1) (N - 2)}{2} & \mbox{if $b = 2 $, } \\
1 & \mbox{if $b = N - 2 - p(N) - 2 r$.} \end{array} \right . \]
Of course, if $N - 2 - p(N) - 2 r = 0$, then $frq (0, E_{r}) \, = \, N - 1$ and if
$ N - 2 - p(N) - 2 r = 2$, then $frq (2, E_{r}) \, = \, 1 + (1/ 2)(N - 1) (N - 2)$.

\item Consider the case where $N\geq4$ is even and $m = N / 2$. Then
\[ {\frak G}(N - 2 m + 2, m, X) \,  =  \, {\frak G}(2, m, X) \, = \,
\frac{(1 - X^{m + 1}) (1 - X^{m + 2})}{(1 - X^{2})}. \]
It is straightforward to verify that
\[\Lambda (N, m) \, = \, \left \{m - 2 r \, \mid \, 0 \leq r \leq \frac{m}{2} \right \}. \]
For $0 \leq r \leq m / 2$, let $A_{r}$ be the $m \times m$ symmetric matrix $[a_{ij}]$, where
\[ a_{ij} \, := \, \left \{\begin{array}{ll} 2 & \mbox{if $i = j$,} \\
 0 & \mbox{if $ \{i, j \} = \{ 2 s - 1, 2 s \}$ for some $0 \leq s \leq r$, } \\
 1 & \mbox{otherwise.} \end{array} \right . \]
Now define $E_{r} \in E(N)$ by setting
\[ E_{r} \, := 2 D_{N} - \mat{2}{0 & A_{r} \\ A_{r} & 0} \]
and for each $L := m - 2 r$, let $G_{L} := Symm_{N}(\delta(z, E_{r}))$. Then the polynomial
$G_{L}$ is homogeneous of total degree
\[ \kappa _{G_{L}} \, := \, \frac{N (3 N - 4)}{4} - L \, = \, m (3 m - 2) - m + 2 r, \]
and its $z_{i}$-degree is at most $2 (N - 1) - m = 3 m - 2$ for $1 \leq i \leq N$. As a consequence of
assertion 4 of the Corollary of Theorem 3, we have $G_{L} \neq 0$. Observe that
\[ suppt (E_{r}) \, =  \, \{0, 1, 2 \}, \;\;\; \]
\[ frq (b, E_{r}) \, = \, \left \{\begin{array}{ll} \frac{N}{2} & \mbox{if $b = 0$,} \\
\frac{N^{2} - 2 N - 8 r }{4} &  \mbox{if $b = 1$,} \\
\frac{N^{2} - 2 N + 8 r }{4} &  \mbox{if $b = 2 $.}
\end{array} \right . \]
Consider the special case where $m$ is an even integer. One can show that
\[ {\Bbb M} (m, m, m / 2, r + m (m -1) / 2) \, \neq \, \emptyset \;\;\;\;\mbox{for}\;\;\;\; 0 \leq r \leq m / 2. \]
So, for $0 \leq r \leq m / 2$, pick a $C_{r} \in {\Bbb M} (m, m, m / 2, r + m (m -1) / 2)$ and define
\[ E_{r} \, := \, \mat{2}{2 D_{m} & 2 C_{r} \\ 2 C_{r}^{T} & 2 D_{m}}. \]
Assertion (i) of Theorem 2 then ensures that for $L = m - 2 r$, the polynomial $G_{L} := Symm_{N}(\delta(z, E_{r}))$
is a nonzero homogeneous polynomial of total degree $m (3 m - 2 ) - m + 2 r$ and its $z_{i}$-degree is at most
$2 (N - 1) - m = 3 m - 2$ for $1 \leq i \leq N$. Also, $suppt (E_{r}) = \{0, 2 \}$. More concretely, let
\[ C_{m / 2} \,  := \,  \mbox{cirmat} ( (a_{1}, \dots , a_{m}) ),\;\;\;\; \]
where $a_{i} = 1$ for $1 \leq i \leq m / 2$ and $a_{i} = 0$ otherwise. For $0 \leq r \leq m / 2$, let $C_{r}$ be
obtained from $C_{m/2}$ by replacing any (randomly picked) $m/2 - r$ entries $1$ in $C_{m/2}$ by $0$. To illustrate,
we exhibit a list of possible $2 C_{r}$ when $N = 8$ and $m = 4$.
\[ 2 C_{2} \,:= \,\mat{4}{2 & 2 & 0 & 0 \\ 0 & 2 & 2 & 0 \\ 0 & 0 & 2 & 2 \\ 2 & 0 & 0 & 2}, \;\;\;\;\;\;
2 C_{1} \,:= \,\mat{4}{2 & 2 & 0 & 0 \\ 0 & 2 & 2 & 0 \\ 0 & 0 & 2 & 2 \\ 0 & 0 & 0 & 2}, \]  \\
\[ 2 C_{0} \,:= \,\mat{4}{0 & 2 & 0 & 0 \\ 2 & 0 & 2 & 0  \\ 0 & 2 & 0 & 2 \\ 0 & 0 & 2 & 0}.  \]

\item Consider the case where $N \geq 5$ is odd and $m = (N + 1) / 2$. Then
\[ {\frak G}(N - 2 m + 2, m, X) \,  =  \, {\frak G}(1, m, X) \, = \, (1 - X^{m + 1}). \]
Letting $N := 2 n + 1$, we have $m = n + 1$ and
\[\Lambda (N, m) \, = \, \left \{ \frac{N + 1}{4} \right \} \, = \, \left \{ \frac{n + 1}{2} \right \} . \]
Let $A$ be the $n \times (n + 1)$ matrix $[a_{ij}]$ such that for $1 \leq i \leq n$ and $1 \leq j \leq n + 1$,
\[ a_{ij} \, := \, \left \{\begin{array}{ll} 1 & \mbox{if $i \neq j$ and $(i, j) \neq (n, n + 1)$,}\\
2 & \mbox{otherwise.} \end{array} \right . \]
Let $E \in E(N)$ be the matrix defined in block-form by
\[ E\, := \, 2 D_{N} - \mat{2}{0 & A  \\ A^{T} & 0}. \]
Recalling the definitions preceding Theorem 3, it is easily verified in this case that
\[ grp (A) \, = \, \{ \theta \in S_{n} \, \mid \, \theta (n) = n \} \]
and $rat (A, T) = \{ pol (A, T) ^{-1} \}$, where
\[ pol (A, T) \, = \, \prod_{r=1}^{n - 1} (T_{r} - T_{n}) \prod_{1 \leq r < s \leq n} (T_{r} - T_{s})^{2}. \]
As a consequence, $A$ is seen to be an admissible matrix. Now Theorem 3 allows us to conclude that
$G := Symm_{N}(\delta(z, E))$ is a nonzero polynomial which is homogeneous of total degree
\[\kappa _{G} \, := \, \frac{N (3 N - 5)}{4} - \frac{(N + 1)}{4} \, = \, N (N -1) - \frac{(N + 1)^2}{4},  \]
and its $z_{i}$-degree is at most $2 (N - 1) - m = 3 n - 1$ for $1 \leq i \leq N$. We have
\[ suppt (E) \, =  \, \{0, 1, 2 \}, \;\;\; \]
\[ frq (b, E) \, = \, \left \{\begin{array}{ll} \frac{N + 1}{2}& \mbox{if $b = 0$,}\\
\frac{(N - 3)(N + 1)}{4} & \mbox{if $b = 1$,} \\
\frac{(N - 1)^2}{4} & \mbox{if $b = 2$.} \end{array} \right . \]

\item Consider the case where $N$ is even and $m = 1 + (N/2)$. Since
$\Lambda (N, 1 + (N/2)) = \{0 \}$, we must have $L = 0$. Let $P(N)$ denote the $N \times N$ symmetric matrix whose
entries $[p (i, j)]$ are defined as follows: assuming $(i, j) = (l_{1}(N/2) + r_{1}, l_{2}(N/2) + r_{2})$,
where $l_{1}, l_{2}\in\{0,1\}$ and $1 \leq r_{1} , r_{2} \leq N/2$,
\[ p(i,j) \, := \, \left \{ \begin{array}{ll} 2  & \mbox{if $l_{1} = l_{2}$ and $r_{1} \neq r_{2}$,} \\ 0 &
\mbox{if $r_{1} = r_{2}$, } \\ 1 & \mbox{otherwise.} \end{array} \right . \] \\Let $G := Symm_{N}(\delta(z, E))$,
where $E:=P(N)$.  Then $G$ is nonzero by the Corollary to \cite[Theorem 3]{MQS}. For even $m$, we also have the option of letting $G := Symm_{N}(\delta(z, E_{0}))$,
where $E_{0} := M_{0}(N / 2, 2, 1, 1)$, which is nonzero by assertion (ii) of Theorem 2.  If $N = 4$, then since the space of binary invariants of type $(4, 3)$ has dimension $1$, our $G$ is  essentially ({\it i.e.}, up to numerical multiples) the only nonzero binary invariant of type $(4, 3)$.

\item For arbitrary values of $N$ and $m$, it is not possible to obtain an explicit listing of the set
$\Lambda (N, m)$. Therefore, we shall remain content to consider all possible values of $m$ only when
$N \leq 8$. In view of the cases dealt with above, it only remains to deal with $(N, m) = (7, 3)$ and
$(N, m) = (8, 3)$. Recall that $D_{r, s}$ denotes the $r \times s$ matrix whose $ij$-th entry is
$(1 - \delta_{ij})$, where $\delta _{ij}$ is the {\it Kronecker delta}, and $D_{r} = D_{r,r}$.

\begin{description}
\item[$(N = 7, \, m = 3)$:] In this case $2 (N - 1) - m = 9$ and
\[ \Lambda (7, 3) \,  = \, \left \{ \frac{3}{2},\; \frac{5}{2}, \;\frac{9}{2} \right \}. \]
For $L \in \Lambda (7, 3)$, let $G_{L} := Symm_{N}(\delta(z, A_{L}))$, where $A_{L} \in E(7)$
is defined as follows:
\[ A_{3/2} \, := \, 2 D_{7} - \mat{2}{0 & C \\ C^{T} & 0}, \;\;\;
\mbox{where} \;\;\; C \,:= \,\mat{4}{2 & 0 & 1 & 1 \\ 1 & 1 & 0 & 2 \\ 0 & 2 & 2 & 0} , \]
\[A_{5/2} \, := \, 2 D_{7} - \mat{2}{0 & C \\ C^{T} & 0}, \;\;\;
\mbox{where} \;\;\; C \,:= \,\mat{4}{2 & 1 & 1 & 1 \\ 1 & 1 & 0 & 2 \\ 0 & 2 & 2 & 0}, \]
\[ A_{9/2} \,:= \, 2 D_{7} - \mat{2}{0 & C \\ C^{T} & 0}, \;\;\; \mbox{where} \;\;\;
C \,:= \,\mat{4}{2 & 1 & 1 & 1  \\ 1 & 1 & 1 & 2  \\ 0 & 2 & 2 & 1 }. \]
Then, Theorem 3 ensures that $G_{3/2}$, $G_{5/2}$ and $G_{9/2}$ are nonzero homogeneous
polynomials of total degrees $30$, $29$ and $27$, respectively. Moreover, the $z_{i}$-degree
of each $G_{L}$ is at most $9$ for $1 \leq i \leq 7$.

\item[$(N = 8, \, m = 3)$:] Now $2 (N - 1) - m = 11$ and
\[ \Lambda (8, 3) \, = \, \{ 0, 2, 3, 4, 6 \}. \]
For $L \in \Lambda (8, 3)$, let $G_{L} := Symm_{N}(\delta(z, A_{L}))$, where $A_{L} \in E(8)$
is defined as follows:
\[ A_{0} \, := \, \mat{2}{3 D_{3} & C \\ C^{T} & 2 D_{5}}, \;\;\;
\mbox{where} \;\;\; C \,:= \,\mat{5}{2 & 0 & 1 & 2 & 0 \\ 1 & 2 & 0 & 1 & 1 \\ 0 & 1 & 2 & 0 & 2 }, \]
\[ A_{2} \, := \, \mat{2}{3 D_{3} & C \\ C^{T} & 2 D_{5}}, \;\;\;
\mbox{where} \;\;\; C \,:= \,\mat{5}{2 & 0 & 1 & 2 & 0 \\ 1 & 2 & 0 & 0 & 0 \\ 0 & 1 & 2 & 0 & 2 }, \]
\[ A_{3} \,:= \, 2 D_{8} - \mat{2}{0 & C \\ C^{T} & 0}, \;\;\; \mbox{where} \;\;\;
C \,:= \,\mat{5}{0 & 2 & 1 & 0 & 2 \\ 2 & 1 & 0 & 2 & 1 \\ 1 & 0 & 2 & 1 & 0}, \]
\[ A_{4} \,:= \, 2 D_{8} - \mat{2}{0 & C \\ C^{T} & 0}, \;\;\; \mbox{where}\;\;\;
C \,:= \,\mat{5}{2 & 2 & 1 & 0 & 1 \\0 & 1 & 2 & 1 & 1 \\1 & 0 & 1 & 2 & 1}, \]
\[ A_{6} \, := \, \mat{2}{2 D_{3} & D_{3, 5} \\ D_{5,3} & 2 D_{5}} \;\;\mbox{or} \;\; M(0 < 3 < 8, 1, 1). \]
A SAGE computation (thanks to Luis Finotti) shows that $G_{0}$, $G_{2}$ are nonzero; in fact, their evaluations
at $z_{i} = i - 1$ for $1 \leq i \leq 8$ are nonzero integers. Next, Theorem 3 ensures that
$G_{3}$ and $G_{4}$ are nonzero. Lastly, assertion (ii) of \cite[Theorem 3]{MQS} ensures that $G_{6} \neq 0$.
Here, for each $L$, $G_{L}$ is homogeneous of total degree $44 - L$ and its $z_{i}$-degree does not exceed $11$
for $1 \leq i \leq 8$. \\
\end{description}
\end{enumerate}

Combining the results of items (1) through (5) above, along with the calculations in (6), yields Theorem 1 in the introduction. \hfill \qed \\

\noindent \underline{\bf Remarks and Questions}:
\begin{enumerate}
\item Let $E_{r}$ be as in the first part of the case $m = N / 2$ considered above. If $N \geq 12$
and $r \geq 2$, then we do not know whether $Symm_{N} (\delta (z, E_{r}))$ is nonzero. For example,
when $N = 12$ ($m = 6$) and $r = 2$, the corresponding $A_{2}$ is not admissible and hence Theorem 3
cannot be applied. So, the open questions: for what values of $2 \leq r \leq m / 2$ is (1) $A_{r}$
admissible and (2) $Symm_{N} (\delta (z, E_{r}))$ nonzero?

\item For the choice of $A_{0}$, $A_{2}$ in the above $(N, m) = (8, 3)$ case,
none of our theorems seem to ensure that the corresponding $G_{0}$, $G_{2}$ are nonzero and hence
we are forced to be content with merely a computational verification.  Furthermore, it is seen that for any choice of $A_0$, $A_2$, at least one of the entries has to be $\geq$ 3.

\item In the case of $(N, m) = (8, 3)$, disregarding the requirement of $2 D_{5}$ as a diagonal block
leads to further choices for $G_2$, $G_4$ and $G_6$:
\[A_{2} \, := \, 2 D_{8} - \mat{2}{0 & C \\ C & 0},\;\;\;
\mbox{where} \;\;\; C \,:= \,\mat{4}{2 & 1 & 0 & 0 \\ 1 & 2 & 1 & 0 \\ 0 & 1 & 2 & 1 \\ 0 & 0 & 1 & 2}. \]
Then Theorem 3 ensures that $G_{2}$ is nonzero.
\[\begin{array}{c} A_{4} \,:= \, \mat{2}{B & C \\ C^{T} & 2 D_{5}}, \;\;\; \mbox{where}\;\;\;
 B \,:= \,\mat{3}{0 & 2 & 4 \\ 2 & 0 & 4 \\ 4 & 4 & 0}\;\;\;\mbox{and} \\  \\
C \,:= \,\mat{5}{2 & 0 & 0 & 2 & 0 \\0 & 2 & 0 & 0 & 2 \\0 & 0 & 2 & 0 & 0}. \end{array} \]
Then assertion (i) of Theorem 2 ensures that $G_{4}$ is nonzero.
\[ A_{6} \, := \, \mat{2}{0 & C \\ C^{T} & 0}, \;\;\; \mbox{where} \;\;\;
C \,:= \, \mat{4}{2 & 2 & 2 & 4 \\ 2 & 2 & 2 & 4 \\ 2 & 2 & 2 & 2 \\ 4 & 4 & 2 & 0}. \]
Then assertion (i) of Theorem 2 ensures that $G_{6}$ is nonzero.

\item Given integers $N$, $m$ with $N \geq 3$, $3 \leq m \leq 1 + (N/2)$ and
given a half-integer $L \in \Lambda (N, m)$, what restrictions on $(N, m, L)$ are necessary and
sufficient for there to exist an $A \in E(N, \leq 2 (N - 1) - m)$ such that $A$ has $2 D_{N - m}$
as a diagonal block, with $\| A \| = N d - 2 L$ and $Symm_{N}(\delta(z, A))$ nonzero? \\
\end{enumerate}

\bibliographystyle{plain}
\bibliography{zFermMQS3}
\end{document}